\documentclass[12pt]{article}

\def\Title{Integration of Superforms and Super-Thom Class}
\def\titlewidth{15cm}
\def\Author{Pietro A. Grassi and Matteo Marescotti}
\def\Date{December, 2007}
\def\Subject{}
\def\Keywords{}

\usepackage{amsmath,amsfonts,amssymb,latexsym,amscd,marvosym}
\usepackage[colorlinks=false,hyperfootnotes=false,urlbordercolor={0 0 1},citebordercolor={0 1 0},linkbordercolor={1 0 0},linktocpage=true,pdftitle={\Title},pdfauthor={\Author},pdfsubject={\Subject},pdfkeywords={\Keywords},bookmarksopen=true,plainpages=false,breaklinks=true]{hyperref}

\def\hhref#1{{\tt[\href{http://arxiv.org/abs/#1}{#1}]}}
\def\references{
\vskip 0.2cm
\vskip \baselineskip
\addcontentsline{toc}{section}{References}
\phantom{?}\vspace{-2\baselineskip}
}

\makeatletter
\def\tableofcontents{%
  \section*{\contentsname}%
\vskip 0.2cm
  \@starttoc{toc}\vskip 1.2cm}
\renewcommand\l@section[2]{%
  \ifnum \c@tocdepth >\z@
    \addpenalty\@secpenalty
    \addvspace{1.0em \@plus\p@}%
    \setlength\@tempdima{1.5em}%
    \begingroup
      \parindent \z@ \rightskip \@pnumwidth
      \parfillskip -\@pnumwidth
      \leavevmode {\bfseries
      \advance\leftskip\@tempdima
      #1}\nobreak\
      \leaders\hbox{$\m@th\mkern \@dotsep mu\hbox{.}\mkern \@dotsep mu$}
     \hfil \nobreak\hb@xt@\@pnumwidth{\hss\bfseries #2}\par
    \endgroup
  \fi}

\renewcommand{\section}{\@startsection
  {section}%                % the name
  {1}%                      % the level
  {0mm}%                    % the indent
  {-\baselineskip}%         % the beforeskip
  {0.75\baselineskip}%       % the afterskip
  {\bfseries\Large}%        % the style
}%

\renewcommand{\subsection}{\@startsection
  {subsection}%
  {2}%
  {0mm}%
  {-\baselineskip}%
  {0.5\baselineskip}%
  {\bfseries\large}%
}%

\renewcommand{\subsubsection}{\@startsection
  {subsubsection}%
  {3}%
  {0mm}%
  {-\baselineskip}%
  {0.5\baselineskip}%       %{-1em}%
  {\bfseries\normalsize}%
}%

\long\def\@fnsymbol#1{\ifcase#1\hbox{}\or *\or \Football \or \Coffeecup \or {\footnotesize\Bicycle}\else\@ctrerr\fi\relax}
\makeatother

\numberwithin{equation}{section}
\long\def\symbolfootnote[#1]#2{\begingroup%
\def\thefootnote{\fnsymbol{footnote}}\footnote[#1]{#2}\endgroup}

\def\a{\alpha}
\def\b{\beta}
\def\g{\gamma}
\def\l{\lambda}
\def\m{\mu}
\def\n{\nu}
\def\d{\delta}
\def\e{\epsilon}
\def\t{\theta}
\def\r{\rho}
\def\s{\sigma}

\def\M{\mathcal{M}}

\def\G{\mathcal{G}}
\def\mfg{\mathfrak{g}}
\def\E{\mathcal{E}}
\def\B{\mathcal{B}}
\def\W{\mathcal{W}}
\def\P{\mathcal{P}}
\def\H{\mathcal{H}}
\def\L{\Lambda}

\def\half{\frac{1}{2}}
\def\de{\mathrm{d}}
\def\lie{\mathcal{L}}
\def\plie{\hat{\lie}}
\def\intd{\mathrm{i}}
\def\co{e}
\def\SO{\text{SO}}

\renewcommand{\varOmega}{\Omega}
\newcommand{\sdet}{\text{sdet\,}}
\newcommand{\eqdef}{\stackrel{\text{\tiny def}}{=}}
\newcommand{\be}{\begin{equation}}
\newcommand{\ee}{\end{equation}}

\begin{document}

%%%%%%%%%%%%%%%%% Title Page %%%%%%%%%%%%%%%%%%%%%%%%%%%%%%%%%%
\thispagestyle{empty}
\rightline{\tt \href{http://arXiv.org/abs/hep-th/yymmnnn}{hep-th/yymmnnn}}
\rightline{\tt DISTA-UPO-07}\vskip -0.3mm
\rightline{\tt \Date}
\vskip 0.9cm\normalsize
\hrule
\vskip 1.2cm\noindent\quad
\parbox[h]{\titlewidth}{\bfseries\Large\Title}
\vskip 1.2cm\noindent
{\quad\bfseries Pietro Antonio Grassi}$\,\symbolfootnote[4]{\,\tt \href{mailto:pgrassi@mfn.unipmn.it}{pgrassi@mfn.unipmn.it}}$ {\bfseries and Matteo Marescotti}\,\symbolfootnote[2]{\,\tt \href{mailto:marescot@mfn.unipmn.it}{marescot@mfn.unipmn.it}}
\vskip .3 truecm
\small\noindent
\quad \hspace{0.4mm}DISTA, Universit\`a del Piemonte Orientale, Via Bellini 25/{\scriptsize G}\,, I-15100, Alessandria, Italy
\vskip .03cm\noindent
\quad \hspace{0.4mm}INFN, sezione di Torino, gruppo collegato di Alessandria, Italy\\
\vskip 0.4cm\normalsize
\hrule\vskip 2.8cm
\subsubsection*{Abstract}

We address the basic problem of constructing the Thom class for a supermanifold. 
Given a cohomological class of a supermanifold and the restriction of the supermanifold 
to its bosonic submanifold, the Thom class gives a prescription to define the integral over the 
bosonic submanifold in terms of the integral over the entire supermanifold. In addition, 
we provide some new 
interesting examples of supermanifolds obtained by extending a given bosonic manifold, 
we discuss the construction of superforms of special type (which transform as Berezinian under 
change of supercoordinates) and we define the de Rham cohomology. We review 
the construction of the Thom class in the conventional geometry and we discuss the extension 
to the supermanifolds. Then, we compute explicitly the Thom class 
for the case of ${\mathbb {CP}}^{(1|2)}$ and, as expected, the result is singular. We provide a 
regularization technique to handle the fermionic Thom class in practical applications. We conclude with 
some remarks about Calabi-Yau spaces and their embedding into super-CY.   

\newpage
\tableofcontents

%%%%%%%%%%%%%%%%%%%%%%%%% Introduction %%%%%%%%%%%%%%%%%%%%%%%%%%%%
\section{Introduction}

Supermanifolds are an interesting  mathematical constructions that
have their roots in the formulation of supersymmetry and supergravity. They
can be thought as bosonic manifolds with anticommuting "fibers". In reality, they are better
constructed in terms of sheaf theory (see for example \cite{Manin:1988ds,BBH,vara}) where the sheaf over a given
open set of the bosonic manifold underneath is a Grasmann algebra with even and odd generators. They
become sets of points by means of the functor-of-point construction. The geometry of these ringed
spaces has been largely studied in \cite{Manin:1988ds} and in \cite{BBH}. A more pedagogical account with applications to superprojective spaces has been published in year 2007 \cite{catenacci}. We refer to this work for more details on superprojective spaces.

Recently, the use of supermanifolds in string theory \cite{Berkovits:2000fe} and twistor string theory
\cite{Witten:2003nn} has revealed new aspects of their geometry that are worth investigating. One of these aspects is the possibility to construct super-Calabi-Yau manifolds (supermanifolds which are K\"ahler and super-Ricci flat  \cite{Rocek:2004bi,Zhou:2004su,Rocek:2004ha,Belhaj:2004ts})
where topological strings can live and propagate \cite{Neitzke:2004pf,Aganagic:2004yh,Popov:2004rb,Kumar:2004dj,Popov:2004nk,Ahn:2004xs}. Sigma models on supermanifolds have been considered several
times in the literature and we have to mention the work \cite{Konechny:1997hr,Schwarz:1995ak} where
super-Calabi-Yau spaces have been already taken into account. 
The super-CY spaces 
were introduced in string theory in paper \cite{Sethi:1994ch}. There the sigma models with supertarget spaces, their conformal invariance and the analysis of some topological rings were studied. 
Recently new studies  on sigma models with supercoset target spaces have been published \cite{mt,berADS,Kagan:2005wt,bervafa}.

As is been shown by Witten in \cite{Witten:2003nn}, given a super-Ricci K\"ahler manifold, one can build a topological B-model twisting a
$N=2$ superconformal sigma model. The resulting model has vanishing conformal
charge and the correlation functions among vertex operators can be defined at any genus of the underlying
worldsheet. The advantage of this  is
the computation of topological invariants of the super-Calabi-Yau
space in terms of correlation functions of topological strings
\cite{Witten:1991zz,Hori:2003ic}. This is not a novelty for
topological strings: indeed, in ``bosonic'' cases the interesting spaces
are Calabi-Yau three folds for which the topological invariants are computed
for deriving the low-energy coupling constants. However, usually the
classical techniques of evaluating those invariants requires a complete
knowledge of the functional expression and a method to integrate over
the Calabi-Yau space. Fortunately, one can avoid the direct computation
of those invariants, by computing the correlation functions of the topological
strings living on these Calabi-Yau spaces.

Let us recall that a Calabi-Yau is a complex K\"ahler manifold with
vanishing first Chern class. For those manifold, one can construct a B
model of twisted N=2 superconformal symmetry. The resulting theory
is much richer then the original classical space parametrized by the
bosonic coordinates; nevertheless, the path integral for the
quantum model is exact namely the quantum corrections
do not affect the computing the classical invariants (different story for
A-models where worldsheet instantons affect the result). It turns out that
the quantum path integral is easier to compute since it does not
require a complete information about the functional form
of a given topological invariants (whose integral is a number). Obviously,
there should be a way to show that the geometrical computation and
the quantum formulation are equivalent. This means that the
integral of the topological invariants should be expressed as integrals in a
larger space with a precise prescription. This
is known as {\it localization} technique and it is based on the equivariant cohomology
(see \cite{Cordes:1994fc} for a
pedagogical accounts with several references to the original literature).

The localization techniques uses the fact that it is usually
easier to compute the integral of a form on a larger space than the
original one, viewed as a submanifold. By specifying the Thom class (for
example in the case of a hypersurface in a toric variety), the integration is
performed in the total variety localizing the integrals on the hypersurface.
The Thom class is a fancy way to defined a Dirac delta function projecting
the integral on the larger space down to the original space. One can
see how to construct the Thom class and the Thom homomorphism in
\cite{Bott:1997fr}. The Thom class has its origin in differential geometry
and equivariant cohomological theories, but it can be easily represented
by means of quantum field theory techniques (see \cite{Cordes:1994fc}). 
Here we repeat the same construction by embedding a bosonic manifold into a 
supermanifold. 

The basic problem we address here is the following: given a cohomological class 
$\xi$ belonging to the group $H^{\bullet}(\hat \M)$ of the supermanifold $\hat{\M}$ and the inclusion $i: \M \rightarrow \hat \M$ of the bosonic submanifold $\M$ into $\hat\M$, 
the corresponding Thom class $\eta[\M \rightarrow \hat \M]$ is defined as follows 
\begin{equation*}
\int_\M i^*(\xi) = \int_{\hat \M} \xi \wedge \eta[\M \rightarrow \hat \M]\,.
\end{equation*}
 with $\eta[\M \rightarrow \hat \M]$ belonging to the space $H^{\bullet}(\hat \M)$ as will be discussed later. Of course the integration of a superform must be clarified. That leads to an interesting geometrical construction \cite{Belopolsky:1997jz}.

We mention the work of Bruzzo and Fucito \cite{Bruzzo:2003rw} where
the localization technique has been extended to superinstantons and
embedded in supergeometry in order to take into account the supermoduli. In
the work of Lavaud \cite{lavaud} a further refinement of the localization idea has been
proposed and the construction of equivariant cohomology is studied. We use a
different approach based on a quantum field theory approach in order
to create a bridge between the mathematical theory of localization and more physical applications.
In particular, we start from a given bosonic manifold ${\cal M}$ and
we construct on it a supermanifold $\widehat{\cal M}$ which is a super-Calabi-Yau (the original space ${\cal M}$ must be at least complex and K\"ahler,
and we follow \cite{Grassi:2006cd} for the construction).
Then, we construct the Thom class for reducing the integral over the supermanifold $\widetilde{\cal M}$ to the integral on  ${\cal M}$. We have to recall that the space of differential forms in the case of supermanifolds has
a important difference with the usual ``bosonic differential geometry''.
As is been discussed in \cite{Grassi:2004tv}, one needs a special type of differential forms to define integral of superforms. We develop further
this theory, by constructing the mapping of cohomologies with special
operators known as {\it picture changing operators}
(see \cite{Friedan:1985ey} for the definition of these operators in fermionic
strings and \cite{Berkovits:2004px,Grassi:2004tv} for picture changing operators in Pure
Spinor formalism). In particular, we show
how given a form in ${\cal M}$ to construct the corresponding form
in $\hat{\cal M}$  has to be integrated. We construct the Thom class  and
we shown that the latter coincides with the picture changing operators
constructed in \cite{Berkovits:2004px}. In addition, we find also
the representation used in \cite{Berkovits:2006vi}.

We consider an application for the localization technique to
define the integration over a super-Calabi-Yau. The supermanifold
is constructed from a bosonic manifold following \cite{Grassi:2006cd} and
we give for the Thom class. 
%We repeat the analysis for a bosonic
%manifold $AdS_{5} \times S^{5}$ where its superextension
%is $PSU(2,2|4)/SO(1,4) \times SO(5)$.
In order to illustrate the construction we consider the case of embedding 
${\mathbb CP}^{(1)}$ in the SCY ${\mathbb CP}^{(1|2)}$. We select a given cohomological 
class and we construct the Thom class. It will be noticed that the integration over the auxiliary 
fields needed to represent the Thom class using the BRST technique leads to a singularity. However, 
this can be regularized using a second BRST quartet and a suitable gauge fermion.  

The paper is organized as follows: in sec. 2 we present a new derivation of the embedding of 
a bosonic manifold into a supermanifold and we present some equations for the superprojective 
space useful for the other sections. In sec. 3 we construct the differential forms for supermanifolds and 
the supercomplex of singular forms. In sec. 4 we construct the equivaraint fermionic Thom class and 
we provide a useful regularization technique to define the Thom class for path integral computations. Finally, in sec. 5 we present some preliminary results regarding the SCY computations using 
the Thom class. 

%%%%%%%%%%%%%%%%%%%%% Section 1 %%%%%%%%%%%%%%%%%%%%%%%

\section{Supermanifolds}

\subsection{Supermanifolds with K\"ahler-Einstein bosonic submanifolds.}

In order to provide some interesting example of supermanifolds, we refer to the construction 
given in \cite{Grassi:2006cd}. Given a bosonic manifold, we want to study an interesting 
super extension (for example with vanishing first Chern class) and we want to use that 
supermanifold to study the integration.\footnote{An interesting paper exploring the relation between supergravity background and Ricci-flat supermanifolds appeared recently \cite{Tomasiello:2007eq}.}

A similar construction to \cite{Grassi:2006cd} is also possible for even-dimensional Einstein manifolds having complex structures compatible with their metric tensors, i.e. K\"ahler-Einstein manifolds. Let $\mathcal{M}$ be a K\"ahler-Einstein manifold of complex dimension $m$ endowed with the Hermitian metric $g_{m\bar{n}}$\,, and let $K$ be a K\"ahler potential for $g$\,. Consider now a Hermitian supermanifold $\mathcal{\hat{M}}$ of complex dimension $(m|n)$ having $\mathcal{M}$ as a body and equipped with the supermetric 
\begin{equation}\label{Ksuperext}
\begin{gathered}
G_{m\bar{n}}=\frac{g_{m\bar{n}}}{1+\frac{\varLambda}{n}\left|\det g\right|^{\frac{1}{n}}\t\bar{\t}}+\frac{\partial_m\!\left|\det g\right|\,\partial_{\bar{n}}\!\left|\det g\right|\,\,\t\bar{\t}}{n^2\left|\det g\right|^{2-\frac{1}{n}}\left(1+\frac{\varLambda}{n}\left|\det g\right|^{\frac{1}{n}}\t\bar{\t}\right)^{\!2}}\,\,,\\
G_{m\bar{\n}}=i\,\,\frac{\left|\det g\right|^{\frac{1}{n}-1}\partial_m\!\left|\det g\right|}{n\left(1+\frac{\varLambda}{n}\left|\det g\right|^{\frac{1}{n}}\t\bar{\t}\right)^{\!2}}\,\,\t_{\bar{\n}}\,,\\
G_{\m\bar{n}}=i\,\,\frac{\left|\det g\right|^{\frac{1}{n}-1}\partial_{\bar{n}}\!\left|\det g\right|}{n\left(1+\frac{\varLambda}{n}\left|\det g\right|^{\frac{1}{n}}\t\bar{\t}\right)^{\!2}}\,\,\t_\m\,,\\
G_{\m\bar{\n}}=\frac{\left|\det g\right|^{\frac{1}{n}}}{1+\frac{\varLambda}{n}\left|\det g\right|^{\frac{1}{n}}\t\bar{\t}}\left(-\,\d_{\m\bar{\n}}+\frac{\frac{\varLambda}{n}\left|\det g\right|^{\frac{1}{n}}\t_\m\t_{\bar{\n}}}{1+\frac{\varLambda}{n}\left|\det g\right|^{\frac{1}{n}}\t\bar{\t}}\right)\,,
\end{gathered}
\end{equation}
where $\t_\m=i\,\t^{\bar{\m}}$, $\t_{\bar{\m}}=-i\,\t^\m$, and $\t\bar{\t}=i\,\t_\m\t^\m$\,. By direct computation one finds that the Ricci supercurvature of $\mathcal{\hat{M}}$ is proportional to $n-m-1\,$, hence if $n=m+1$ then $\mathcal{\hat{M}}$ is Ricci-flat. Furthermore $\mathcal{\hat{M}}$ is also K\"ahler since it turns out that the supermetric \eqref{Ksuperext} is obtainable from the superpotential
\be\label{Kpotext}
\hat{K}= K+\frac{n}{\varLambda}\,\ln\!\left(1+\frac{\varLambda}{n}\left|\det g\right|^{\frac{1}{n}}\t\bar{\t}\right).
\ee
In conclusion on any K\"ahler-Einstein manifold of complex dimension $m$ one can construct a SCY having virtual complex dimension $-1\,$. Typical examples of K\"ahler-Einstein manifolds are the projective spaces $\mathbb{CP}^{\,m}$ endowed with the metrics generated by the Fubini-Study K\"ahler potentials
\begin{equation}\label{fubinistudy2}
K=\ln(1+z\bar{z})\,.
\end{equation}
These manifolds are known to admit SCY extensions $\mathbb{CP}^{(m|m+1)}$\, equipped with the supermetrics generated by the Fubini-Study superpotentials
\begin{equation}\label{fubinistudy1}
\hat{K}=\ln(1+z\bar{z}+\t\bar{\t})\,.
\end{equation}
The projective superspaces $\,\mathbb{CP}^{(m|m+1)}\,$ are indeed partular cases of our construction, since \eqref{Kpotext} coincides with \eqref{fubinistudy1} when $K$ is given by \eqref{fubinistudy2}.

\subsection{Projective superspaces}

As an interesting and simpler example, we provide some formulas regarding the 
case of ${\mathbb {CP}}^{(1|2)}$. In the following section we are going to compute explicitly 
the Thom class for this example and therefore it is useful to give here all ingredients. In particular, 
for the present case we want to compute the class $\eta[{\mathbb {CP}}^{1} 
\rightarrow {\mathbb {CP}}^{1|2}] \in H^{(0|2)}({\mathbb {CP}}^{(1|2)})$  such that, for a given cohomological class\footnote{The superscript $(p|q)$ for a form will be explained in the next section.} 
$\xi \in H^{(1|0)}({\mathbb {CP}}^{(1|2)})$, we have
\begin{equation}
\int_{{\mathbb {CP}}^{(1}} i^*(\xi) = \int_{{\mathbb CP}^{(1|2}}  \xi \wedge 
\eta[{\mathbb {CP}}^{1} \rightarrow {\mathbb {CP}}^{(1|2)}]\,.
\end{equation}
 The case of ${\mathbb {CP}}^{(1|2)}$ is interesting since is a super-Calabi-Yau, its Ricci 
 flat metric is known and the integration over ${\mathbb {CP}^{(1)}}$ is well-defined. The same 
 ingredients can be computed for any supermanifold of the previous section, but they 
 are rahter cumbersome and not very illustrative.  
 
The Fubini-Study supermetric of a superprojective
$\mathbb{CP}^{(m|n)}$ is given in components by
\begin{equation}
\begin{gathered}
G_{m\bar{n}}=\frac{1}{1+z\bar{z}+\t\bar{\t}}\,\left(\delta_{m\bar{n}}-\frac{z^{\bar{m}}z^n}{1+z\bar{z}+\t\bar{\t}}\right)\,,\\
G_{m\bar{\nu}}=-\frac{z^{\bar{m}}\t^{\nu}}{(1+z\bar{z}+\t\bar{\t})^2}\,,\\
G_{\mu\bar{n}}=\frac{\t^{\bar{\mu}}z^n}{(1+z\bar{z}+\t\bar{\t})^2}\,,\\
G_{\mu\bar{\nu}}=-\frac{1}{1+z\bar{z}+\t\bar{\t}}\,\left(\delta_{\mu\bar{\nu}}-\frac{\t^{\bar{\mu}}\t^\nu}{1+z\bar{z}+\t\bar{\t}}\right)\,.
\end{gathered}
\end{equation}
Therefore the inverse supermetric is
\begin{equation}
\begin{gathered}
G^{\,\bar{m}n}=\left(1+z\bar{z}+\t\bar{\t}\right)\,\left(\delta^{\,\bar{m}n}+z^{\bar{m}}z^n\right)\,,\\
G^{\bar{m}\nu}=\left(1+z\bar{z}+\t\bar{\t}\right)\,z^{\bar{m}}\t^{\nu}\,,\\
G^{\,\bar{\mu}n}=\left(1+z\bar{z}+\t\bar{\t}\right)\,\t^{\bar{\mu}}z^{n}\,,\\
G^{\,\bar{\mu}\nu}=\left(1+z\bar{z}+\t\bar{\t}\right)\,\left(\delta^{\,\bar{\mu}\nu}+\t^{\,\bar{\mu}}\t^{\nu}\right)\,.
\end{gathered}
\end{equation}
The corresponding Levi-Civita superconnection is
\be
\begin{gathered}
\Gamma^{m}_{~np}=-\frac{\delta^m_{~n}z^{\bar{p}}+\delta^m_{~p}z^{\bar{n}}}{1+z\bar{z}+\t\bar{\t}}\,,\qquad
\Gamma^{m}_{~n\rho}=\Gamma^{m}_{~\rho n}=\frac{\delta^m_{~n}\t^{\bar{\rho}}}{1+z\bar{z}+\t\bar{\t}}\,,\qquad \Gamma^m_{~\nu\rho}=0\\
\Gamma^{\mu}_{~np}=0\,,\qquad\Gamma^{\mu}_{~\nu p}=\Gamma^{\mu}_{~p\nu}=-\frac{\delta^\mu_{~\nu}z^{\bar{p}}}{1+z\bar{z}+\t\bar{\t}}\,,\qquad \Gamma^{\mu}_{~\nu\rho}=\frac{\delta^\mu_{~\nu}\t^{\bar{\rho}}-\delta^\mu_{~\rho}\t^{\bar{\nu}}}{1+z\bar{z}+\t\bar{\t}}\,,
\end{gathered}
\ee from which one finds the Riemann supercurvature
\begin{equation}
R^M_{~NP\bar{Q}}=-\,\Gamma^M_{~NP,\bar{Q}}=\delta^M_{~N}\,G_{P\bar{Q}}+(-1)^{NP}\delta^M_{~P}\,G_{N\bar{Q}}\,,
\end{equation}
and Ricci supertensor
\begin{equation}
R_{M\bar{N}}=(d_B-d_F+1)\,\,G_{M\bar{N}}\,.
\end{equation}
In particular the superprojective $\mathbb{CP}^{(1|2)}$ the
supermetric reduces to the form
\begin{equation}
\begin{gathered}
G_{z\bar{z}}=\frac{1+\t\bar{\t}}{(1+z\bar{z}+\t\bar{\t})^2}\,,\\
G_{z\bar{\nu}}=-\frac{\bar{z}\,\t^{\nu}}{(1+z\bar{z}+\t\bar{\t})^2}\,,\\
G_{\mu\bar{z}}=\frac{\t^{\bar{\mu}}z}{(1+z\bar{z}+\t\bar{\t})^2}\,,\\
G_{\mu\bar{\nu}}=-\frac{1}{1+z\bar{z}+\t\bar{\t}}\,\left(\delta_{\mu\bar{\nu}}-\frac{\t^{\bar{\mu}}\t^\nu}{1+z\bar{z}+\t\bar{\t}}\right)\,,
\end{gathered}
\end{equation}
from which one finds the inverse supermetric
\begin{equation}
\begin{gathered}
G^{\,\bar{z}z}=\left(1+z\bar{z}+\t\bar{\t}\right)\,\left(1+z\bar{z}\right)\,,\\
G^{\,\bar{z}\nu}=\left(1+z\bar{z}+\t\bar{\t}\right)\,\bar{z}\,\t^{\nu}\,,\\
G^{\,\bar{\mu}z}=\left(1+z\bar{z}+\t\bar{\t}\right)\,\t^{\bar{\mu}}z\,,\\
G^{\,\bar{\mu}\nu}=\left(1+z\bar{z}+\t\bar{\t}\right)\,\left(\delta^{\,\bar{\mu}\nu}+\t^{\,\bar{\mu}}\t^{\nu}\right)
\end{gathered}
\end{equation}
and the superconnection
\begin{equation}
\begin{gathered}
\Gamma^{z}_{~zz}=-\frac{2\,\bar{z}}{1+z\bar{z}+\t\bar{\t}}\,,\qquad
\Gamma^{z}_{~z\rho}=\Gamma^{z}_{~\rho
z}=\frac{\t^{\bar{\rho}}}{1+z\bar{z}+\t\bar{\t}}\,,\qquad
\Gamma^z_{~\nu\rho}=0\,,\\
\Gamma^{\mu}_{~zz}=0\,,\qquad\Gamma^{\mu}_{~\nu
z}=\Gamma^{\mu}_{~z\nu}=-\frac{\delta^\mu_{~\nu}\,\bar{z}}{1+z\bar{z}+\t\bar{\t}}\,,\qquad
\Gamma^{\mu}_{~\nu\rho}=\frac{\delta^\mu_{~\nu}\,\t^{\bar{\rho}}-\delta^\mu_{~\rho}\,\t^{\bar{\nu}}}{1+z\bar{z}+\t\bar{\t}}\,.
\end{gathered}
\end{equation}

It is easy to check from these equations that the superspace  ${\mathbb CP}^{(1|2)}$ has the Fubini-Study metric which is Ricci flat. 

%%%%%%%%%%%%%%%%%%%%%%%%%%%%%%%%%%%%%%%%%%%%%

\section{Differential forms on supermanifolds}

Most of supergeometry can be obtained straightforwardly by extending
the commuting geometry by means of the rule of signs, but this is not the case of the theory of differential forms on supermanifolds. Indeed the naive notion of ``superforms'' obtainable just by adding a $\mathbb{Z}_2$ grading to the exterior algebra turns out not to be suitable for Berezin integration.

A solution to this problem is to introduce the so-called singular $(r|s)$-forms in the game. They are generalized superforms carrying the usual form degree $r$ together with another degree $s$ called \emph{picture degree}. The sum of the twos will be referred to as the \emph{total degree}.

Possible values for the picture degree are $\,0,\ldots,n$ for a supermanifold $\hat \M$ which is locally homeomorphic to ${\mathbb R}^{(m|n)}$ and where $n$ is the number of anticommuting coordinates. We denote by ${\M}$ the bosonic submanifold ${\hat \M}$. In particular $(r|0)$-forms are the naive superforms mentioned above. Therefore their form degree $r$ must be non-negative. On the other hand, when $0<s<n$ the form degree $r$ may assume any integer value, while with $s=n$ the only restriction is $r\le m$. As a consequence, non-trivial singular $(r|s)$-forms may exist with negative form degree.

The resulting de Rham supercomplex, which will be denoted by $\Omega^{(\cdot|\cdot)}(\hat \M)$, is endowed with an exterior derivative $\de:\Omega^{(r|s)}\longrightarrow \Omega^{(r+1|s)}$ and the cohomology is defined for any degrees $r$ and $s$ compatible with the above prescriptions. Since the differential $\de$ does not change the picture degree of a form, the de Rham supercomplex splits into $m+1$ subcomplexes labelled by  different values of $s$\,. These subcomplexes are called \emph{pictures}. It is possible to define picture changing operators which commute with the exterior derivative and map cohomology classes of the $s$-picture to cohomology classes of the $(s+1)$-picture. Restricting to a certain subset of $\Omega^{(\cdot|\cdot)}$, one can also construct inverse picture changing operators.

In the following we use some notations and derivations  of \cite{Belopolsky:1997jz}. We also 
refer to additional material to \cite{Grassi:2004tv}. 

\subsection{Definition}
Let $\mathcal{M}\,$ be a supermanifold of dimension $(m|n)$\,.
A function \,$\xi$\, of $r$ even and $s$ odd tangent vectors to $\mathcal{M}$
is called an $(r|s)$-form if it satisfies the following pair of conditions:
\begin{subequations}\label{singforms}
\begin{gather}\label{jacobian}
\xi(L\hspace{0.5mm}\mathbf{X})=\text{sdet}(L)\,\xi(\mathbf{X})\,,\quad\forall \,L\in \mathrm{GL}(r|s)\,,\\[1mm]
\label{stokes}
\!\!\left(\frac{\vec{\partial}}{\partial X^M_I}\frac{\vec{\partial}}{\partial X^N_J}+(-1)^{IJ+N(I+J)}\,\frac{\vec{\partial}}{\partial X^M_J}\frac{\vec{\partial}}{\partial X^N_I}\right)\xi(\mathbf{X})=0\,,
\end{gather}
\end{subequations}
where we call $\mathbf{X}$ the entire set of arguments $(X_1,\dots,X_r|\hat{X}_1,\dots,\hat{X}_s)$ of $\xi$ and by $I,J$ two labels running from $1$ to $r|s$\,. As an example let us take a set $\co_{\hspace{0.3mm}\mathbf{X}}\!=\!\left(\co_{X_1},\ldots,\co_{X_r}|\co_{\hat{X}_1},\ldots,\co_{\hat{X}_s}\right)$
of $r$ even and $s$ odd covectors. The function of $\mathbf{X}$ given by the formula
\be\label{canform}
\omega_{\mathbf{X}}(\mathbf{Y})=\sdet\!\left(\co_{X_I}(Y_J)\right)
\ee
is an $(r|s)$-form as it verifies both properties \eqref{singforms}\,. Notice that it is singular since a pole arises whenever one of the odd arguments belongs to the null space of the defining odd covectors.

\subsection{Inner and exterior products}
Starting from a given form, one can produce other solutions of \eqref{singforms} by applying certain operations to it. The first operation maps $\Omega^{(r|s)}$ to  $\Omega^{(r+1|s)}$ by taking an exterior product with a covector $X\,$:
\be
\begin{aligned}
(\co_{X}\wedge\hspace{0.5mm}\,\xi)(X_1,\dots,X_{r+1}|&\hat{X}_1,\dots,\hat{X}_{s})\\&\eqdef(-1)^r\left(e_X(X_{r+1})-(-1)^{XJ}e_X(X_J)\,X_{r+1}^M\,\frac{\vec{\partial}}{\partial X_J^M}\right)\xi(\mathbf{X})\,,
\end{aligned}
\ee
with $J=1,\ldots,r|s\,$. The $(r|s)$-form defined in \eqref{canform} can be rewritten by applying the exterior products with the even defining covectors to a $(0|s)$-form:
\be\label{canformint}
\omega_{\mathbf{X}}=\co_{X_1}\wedge\cdots\,\co_{X_r}\wedge\det\!\big(\co_{\hat{X}_\a}(\,\cdot\,)\big)^{-1},
\ee
with $\a=1,\ldots,s$\,. The exterior products supercommute by definition: this means that two of them anticommute when one is taken with an even covector, while they commute when they are both taken with odd covectors. In a compact form one may write the supercommutation rule
\be
\left[\co_X\wedge\,,\co_Y\wedge\right]=0\,.
\ee
An interesting fact is that exterior products with odd vectors can be iterated indefinitely, hence the even degree of the resulting forms is not bounded by the even dimension of the supermanifold. This is a 
well-known feature of superforms.  

The second operation decreases the degree of forms by substituting a given vector for one of its even arguments:
\be\label{innerpr}
(\intd_{X}\hspace{0.2mm}\xi)(X_1,\dots,X_{r-1}|\hat{X}_1,\dots,\hat{X}_s)\eqdef(-1)^{X\xi}\,\xi(X,X_1,\dots,X_{r-1}|\hat{X}_1,\dots,\hat{X}_s)
\ee
For this operation, which is called the inner product with the vector $X$\,, one can make considerations similar to the case of the exterior product. Indeed inner products supercommute by definition: this means again that two of them anticommute when one is taken with an even vector, while they commute when they are both taken with odd vectors. In a compact form one may write the supercommutator
\be
\left[\intd_X,\intd_Y\right]=0
\ee
Inner products with odd vectors can be iterated. So $(\intd_{\hat{X}})^k$ is well defined for differential forms of degree $r\ge k$ and it is non zero. In conclusion exterior and inner products on a supermanifold of dimension $(m|n)$ realize the $\text{OSp}(m\hspace{0.5mm},m|n\hspace{0.5mm},n)$ Clifford superalgebra
\be
\left[\co_X\wedge\,,\co_Y\wedge\right]=0\,, \qquad\left[\intd_X,\intd_Y\right]=0\,,\qquad \left[\co_X\wedge\,,\intd_Y\right]=\co_X(Y)\,.
\ee

The operators introduced till now leave the picture degree of forms unchanged. Now we define new operations which do modify it. The first operation, which will be denoted by $\d(\intd_{\hat{X}})$\,, is very similar to the usual inner product $\intd_{\hat{X}}$\,, but the vector $\hat{X}$ is now substituted for one of the odd arguments of the form:
\be\label{dintd}
\left(\d(\intd_{\hat{X}})\xi\right)(X_1,\dots,X_r|\hat{X}_1,\dots,\hat{X}_{s-1})\eqdef(-1)^r\,\xi(X_1,\dots,X_r|\hat{X},\hat{X}_1,\dots,\hat{X}_{s-1})\,.
\ee
As a consequence $\d(\intd_{\hat{X}})$ maps $\Omega^{(r|s)}$ to $\Omega^{(r|s-1)}$\,. The second operation, which will be denoted by $\d(\co_{\hat{X}}\wedge)$, is defined by the formula
\be\label{dextd}
\begin{aligned}
&\left(\d(\co_{\hat{X}}\wedge)\xi\right)(X_1,\dots,X_r|\hat{X}_1,\dots,\hat{X}_{s+1})\\&\qquad\qquad\qquad\eqdef\frac{1}{\co_{\hat{X}}(\hat{X}_{s+1})}\,\,\xi\left(\dots,X_J-\frac{\co_{\hat{X}}(X_J)\,\hat{X}_{s+1}}{\co_{\hat{X}}(\hat{X}_{s+1})}\,,\dots\right),
\end{aligned}
\ee
with $J=1,\ldots,r|s\,$. It resembles the ordinary exterior product as it increases the total degree of forms by one. The difference is that now the form degree remain unchanged while the picture degree gets modified.

The following identities, which can be easily derived from definitions, mimic a basic property of the $\d$-function and partially justify the notation adopted for operators \eqref{dintd} and \eqref{dextd}\,:
\begin{equation}\label{mimicd}
\begin{gathered}
\d(\intd_{\hat{X}})\,\intd_{\hat{X}}=\intd_{\hat{X}}\,\d(\intd_{\hat{X}})=0\,,\\
\d(\co_{\hat{X}}\wedge)\,\co_{\hat{X}}\wedge=\co_{\hat{X}}\!\wedge\,\d(\co_{\hat{X}}\wedge)=0\,.
\end{gathered}
\end{equation}
It is also important to observe that, as it happens with $\d$-functions, products of two $\d(\intd_{\hat{X}})$'s or $\d(\co_{\hat{X}}\wedge)$'s with the same argument lead to divergencies. On the contrary, when two inner or exterior $\d$-products are taken with linearly independent vectors or covectors, they satisfy the anticommutation rules
\begin{equation}\label{oddantid}
\begin{gathered}
\d(\intd_{\hat{X}})\,\d(\intd_{\hat{Y}})=-\,\d(\intd_{\hat{Y}})\,\d(\intd_{\hat{X}})\,,\\
\d(\co_{\hat{X}}\wedge)\,\d(\co_{\hat{Y}}\wedge)=-\,\d(\co_{\hat{Y}}\wedge)\,\d(\co_{\hat{X}}\wedge)\,.
\end{gathered}
\end{equation}
For completeness it is convenient to define $\d(\intd_{X})$ and $\d(\co_{X}\wedge)$ also for even vectors and covectors as we did for the usual inner and exterior products. The most natural choice is to assume $\d(\intd_{X})=\intd_{X}$ and $\d(\co_{X}\wedge)=\co_{X}\wedge\,$, because this permits to treat even and odd vectors on the same ground. Indeed $\intd_X$ and $\co_X\wedge$ fulfill \eqref{mimicd} and \eqref{oddantid}, and it turns out that the $\d$-function identities
\begin{equation}\label{mimicdgen}
\begin{gathered}
\d(\intd_{X})\,\intd_{X}=\intd_{X}\,\d(\intd_{X})=0\,,\\
\d(\co_{X}\wedge)\,\co_{X}\wedge=\co_{X}\!\wedge\,\d(\co_{X}\wedge)=0\,
\end{gathered}
\end{equation}
and the anticommutation rules
\begin{equation}\label{oddantidgen}
\begin{gathered}
\left\{\d(\intd_{X})\,,\d(\intd_{Y})\right\}=0\,,\\
\left\{\d(\co_{X}\wedge)\,,\d(\co_{Y}\wedge)\right\}=0\,
\end{gathered}
\end{equation}
hold no matter what parity $X$ and $Y$ have. By means of the new operators it is possible to simplify further the expression in \eqref{canformint}. Indeed one can obtain the $(0|s)$-form $\det\!\big(\co_{\hat{X}_\a}(\,\cdot\,)\big)^{-1}$ by applying $\d$-exterior products iteratively to a zero-form:
\be
\det\big(\co_{\hat{X}_\a}(\,\cdot\,)\big)^{-1}=\d(\co_{\hat{X}_1}\wedge)\,\cdots\,\d(\co_{\hat{X}_s}\wedge)\,\,1\,.
\ee
Hence the singular form \eqref{canform} may be rewritten as
\be
\omega_{\mathbf{X}}=\prod_{J=1}^{r|s}\,\d(\co_{X_J}\wedge)\,1\,.
\ee
This shows that $\omega_{\mathbf{X}}$ bears the same relation to supermanifolds of dimension $(r|s)$  as the ordinary form $\co_{X_1}\wedge\ldots\co_{X_r}\wedge\,1=\det(\co_{X_j}(\,\cdot\,))$ bears to $r$-dimensional manifolds. The singularity simply comes from the fact that $\omega_{\mathbf{X}}$ transforms as a Berezinian instead of a usual Jacobian.

Operators $\d(\co_{\hat{X}}\wedge)$ and $\d(\intd_{\hat{Y}})$ do not satisfy the same commutation relations as $\co_{\hat{X}}\wedge$ and $\intd_{\hat{Y}}$\,. Actually their commutators cannot even be defined except for the trivial case $\co_{\hat{X}}\big(\hat{Y}\big)=0$ when they anticommute. However if we consider the weaker form of the usual relation $\left[\co_{\hat{X}}\wedge\,,\intd_{\hat{Y}}\right]=\co_{\hat{X}}\big(\hat{Y}\big)$ which states that
\begin{equation}
\begin{gathered}
\!\co_{\hat{X}}\wedge\intd_{\hat{Y}}\,\xi=\co_{\hat{X}}\big(\hat{Y}\big)\,\xi\qquad \text{if \,}\co_{\hat{X}}\wedge\,\xi=0\,,\\
\intd_{\hat{X}}\,\co_{\hat{Y}}\wedge\,\xi=-\,\co_{\hat{Y}}\big(\hat{X}\big)\,\xi\qquad \text{if \,}\intd_{\hat{X}}\,\xi=0\,,
\end{gathered}
\end{equation}
then there is an analog for $\d(\co_{\hat{X}}\wedge)$ and $\d(\intd_{\hat{Y}})$\,:
\begin{equation}
\begin{gathered}
\d(\co_{\hat{X}}\wedge)\,\d(\intd_{\hat{Y}})\,\xi=\frac{1}{\co_{\hat{X}}\big(\hat{Y}\big)}\,\,\xi\qquad \text{if \,}\co_{\hat{X}}\wedge\,\xi=0\,,\\
\d(\intd_{\hat{X}})\,\d(\co_{\hat{Y}}\wedge)\,\xi=-\,\frac{1}{\co_{\hat{Y}}\big({\hat{X}}\big)}\,\,\xi\qquad \text{if \,}\intd_{\hat{X}}\,\xi=0\,.
\end{gathered}
\end{equation}
The appearence of the contraction $\co_{\hat{X}}\big(\hat{Y}\big)$ in the denominator is consistent with the fact that the $\d$-function is homogeneous of degree $-1$\,.

As one can check from definitions, anticommutators of $\co_{\hat{X}}\wedge$ with $\d(\intd_{\hat{Y}})$ and $\intd_{\hat{X}}$ with $\d(\co_{\hat{Y}}\wedge)$ are always well defined and proportional to $\co_{\hat{X}}\big(\hat{Y}\big)$ with the operatorial part depending only on the argument of the $\d$-functions. Then is possible to define a pair of new operators, which will be denoted by $\d'(\intd_{\hat{X}})$ and $\d'(\co_{\hat{X}}\wedge)$\,:
\begin{equation}
\begin{gathered}
\left\{\co_{\hat{X}}\wedge\,,\d(\intd_{\hat{Y}})\right\}\eqdef\co_{\hat{X}}\big({\hat{Y}}\big)\,\d'(\intd_{\hat{Y}})\,,\\
\left\{\intd_{\hat{X}}\,,\d(\co_{\hat{Y}}\wedge)\right\}\eqdef-\,\co_{\hat{Y}}\big({\hat{X}}\big)\,\d'(\co_{\hat{Y}}\wedge)\,.
\end{gathered}
\end{equation}
Iterating the mechanism one can also define operators with multiple derivatives of $\d$-functions:
\begin{equation}
\begin{gathered}
\left\{\co_{\hat{X}}\wedge\,,\d^{(n)}(\intd_{\hat{Y}})\right\}\eqdef\co_{\hat{X}}\big({\hat{Y}}\big)\,\d^{(n+1)}(\intd_{\hat{Y}})\,,\\
\!\left\{\intd_{\hat{X}}\,,\d^{(n)}(\co_{\hat{Y}}\wedge)\right\}\eqdef-\,\co_{\hat{Y}}\big({\hat{X}}\big)\,\d^{(n+1)}(\co_{\hat{Y}}\wedge)\,.
\end{gathered}
\end{equation}
These operators are very interesting from a cohomological point of view because, as one can check from definitions, their form degrees are
\begin{equation}
\begin{gathered}
\text{deg}\left(\d^{(n)}(\intd_{\hat{Y}})\right)=(n\,|\!-1)\,,\\
\text{deg}\left(\d^{(n)}(\co_{\hat{Y}}\wedge)\right)=(-n\,|1)\,,
\end{gathered}
\end{equation}
hence they can be used to build differential forms with negative degree. To make an example consider the $(0|s)$-form $\d(\co_{\hat{X}_1}\wedge)\cdots\,\d(\co_{\hat{X}_s}\wedge)$\,. By applying $\intd_{\hat{X}}$\,, one gets derivatives of the $\d$-functions and the result is a $(-1|s)$-form:
\be
\intd_{\hat{X}}\,\d\big(\co_{\hat{X}_1}\!\wedge\!\big)\cdots\,\d\big(\co_{\hat{X}_s}\!\wedge\!\big)=\sum_{\a=1}^{s}\,(-1)^\a\,\co_{\hat{X_\a}}\!\big(\hat{X}\big)\,\d\big(\co_{\hat{X}_1}\!\wedge\!\big)\cdots\,\d'\big(\co_{\hat{X}_\a}\!\wedge\!\big)\cdots\,\d\big(\co_{\hat{X}_s}\!\wedge\!\big)
\ee
Recalling that for even vectors and covectors $\d(\intd_{X})$ and $\d(\co_{X}\wedge)$ coincide with $\intd_{X}$ and $\co_{X}\wedge$ respectively, it is natural to assume $\d'(\intd_{X})=1\,,$ $\d'(\co_{X}\wedge)=1\,$ for first derivatives, and $\d^{(n)}(\intd_{X})=0\,,$ $\d^{(n)}(\co_{X}\wedge)=0\,$ for higher derivatives. Indeed with this choice the supercommutation rules
\begin{equation}
\begin{gathered}
\,[\co_X\wedge\,,\d^{(n)}(\intd_Y)]=( Y\cdot\co_X)\,\d^{(n+1)}(\intd_Y)\,,\\
\,[\intd_X\,,\d^{(n)}(\co_Y\wedge)]=(\co_Y\cdot X)\,\d^{(n+1)}(\co_Y\wedge)\,,
\end{gathered}
\end{equation}
hold no matter what parities $X$ and $Y$ have.

\subsection{The differential and the Lie derivative}
A fundamental operator, which maps $\Omega^{(r|s)}$ to $\Omega^{(r+1|s)}$\,, is the de Rham differential $\de\,$. The action of $\de$ can be expressed in coordinates as follows:
\be\label{differential}
\begin{aligned}
(\de\xi)(X;Y_1&,\dots ,Y_{r+1}\,|\,\hat{Y}_1,\dots ,\hat{Y}_s)\\&\eqdef(-1)^r\, Y_{r+1}^M\left(\frac{\vec{\partial}}{\partial X^M}-(-1)^{IM}\,Y_{I}^N\,\frac{\vec{\partial}}{\partial X^N}\frac{\vec{\partial}}{\partial Y^M_I}\right)\xi(X;\mathbf{Y})\,.
\end{aligned}
\ee
By direct computation of the drag of a form produced by a vector field $X$, it easy to find the expression for the Lie derivative $\lie_X$\,. The result is the supercommutator of the differential with the inner product as one expects:
\be\label{deflie}
\lie_X=[\de\,,\intd_X]\,.
\ee
The Lie derivative commutes with the differential,
\be
[\de,\lie_X]=0\,,
\ee
and satisfies the commutation relations
\be\label{comlieintd}
\begin{gathered}
\,[\lie_X\,,\intd_Y]=\intd_{[X,Y]}\,,\\
\,[\lie_X\,,\d^{(n)}(\intd_Y)]=\d^{(n+1)}(\intd_Y)\,\intd_{[X,Y]}
\end{gathered}
\ee
with inner products and their derivatives. The operator $\lie_X$ also appears in the commutator of the differential with $\d^{(n)}\big(\intd_{\hat{X}}\big)$ and the various results can be synthesized in a formula valid for any parity of $X\,$:
\be\label{comddeltaintd}
[\de\,,\d^{(n)}(\intd_X)]=\d^{(n+1)}(\intd_X)\,\lie_X+\frac{1}{2}\,\d^{(n+2)}(\intd_X)\,\intd_{[X,X]}\,.
\ee
For $X$ even, the only nontrivial relation coming from \eqref{comddeltaintd} is the case $n=0\,$ and it is the definition of $\lie_X$ given in \eqref{deflie}. On the other hand for $X$ odd, one obtains nontrivial equations for any value of $n$. Taking $n=0$ leads to the operator $\plie_X$ given by
\be
\plie_X\eqdef[\de\,,\d(\intd_X)]=\d'(\intd_X)\,\lie_X+\frac{1}{2}\,\d''(\intd_X)\,\intd_{[X,X]}\,,
\ee
which is equal to the Lie derivative for $X$ even but it is different for $X$ odd. Indeed $\lie_X$ maps $\Omega^{(r|s)}$ to itself whatever parity $X$ has, while $\plie_X$ maps $\Omega^{(r|s)}$ to itself for $X$ even and $\Omega^{(r|s)}$ to $\Omega^{(r+1|s-1)}$ for $X$ odd. In other words, the total degrees of $\lie_X$ and $\plie_X$ are both zero, but the degree of $\lie_X$ is always $(0|0)$ while the degree of $\plie_X$ for $X$ odd is $(1|-1)\,$. Also $\plie_X$ commutes with the differential,
\be
[\de,\plie_X]=0\,,
\ee
and it satisfies the commutation relation
\be
[\plie_X\,,\d^{(n)}(\intd_Y)]=\d'(\intd_X)\,\d^{(n+1)}(\intd_Y)\,\intd_{[X,Y]}\,,
\ee
from which one finds the analog of \eqref{comlieintd}
\be
[\plie_X\,,\d(\intd_Y)]=\d'(\intd_X)\,\d'(\intd_Y)\,\intd_{[X,Y]}\,.
\ee
With the supercommutators given in the present section one can verify once more the concistency of the $\d$-function notation. Indeed, by adopting the formal integral representation
\begin{equation}
\begin{gathered}
\d(\intd_{\hat{X}})=\frac{1}{2\pi}\int dt~ e^{it\,\intd_{\hat{X}}}\,,\\
\d(\co_{\hat{X}}\wedge)=\frac{1}{2\pi}\int dt~ e^{it\,\co_{\hat{X}}\wedge}
\end{gathered}
\end{equation}
for the inner derivative and exterior product, one can easily reobtain all of the above relations involving $\d$-functions. 

\subsection{The supercomplex}

The de Rham complex of forms on a supermanifold $\mathcal{M}^{(m|n)}$ is a nontrivial extension of the complex of ordinary forms on the body manifold $\mathcal{M}^{(m)}$ (see also 
for a detail analysis of superform complex \cite{Voronov2}).

As we have already pointed out, $(r|0)$-forms are the naive superforms. The corresponding subcomplex starts from $\Omega^{(0|0)}$ on the left but it is unbounded on the right because exterior products $\co_{\hat{X}}\wedge$ commute and they can be iterated indefinitely. So when the picture degree vanishes, the only constraint on the form degree is $r\ge 0\,$.

When $0<s<n$ it is possible again to construct forms with arbitrarily high form degree. Indeed, given the $(0|s)$-form $\d(\co_{\hat{X}_1}\wedge)\cdots\,\d(\co_{\hat{X}_s}\wedge)\,$, one can always find an odd covector $\co_{\hat{X}}$ which is linearly independent from $\co_{\hat{X}_1},\ldots,\co_{\hat{X}_s}$\, since the picture degree $s$ is lower than the odd dimension $n$ of the supermanifold. Then one can apply the exterior product $\co_{\hat{X}}$ indefinitely and generate $(r|s)$-forms with any positive $r$\, as in the case $s=0$\,. But now one can also write $(r|s)$-forms with negative $r$ by using derivatives of $\d$-functions, which do not exist with $s=0$\,. The recipe to get $(-r|s)$-forms is to use $s$ exterior products $\co_{\hat{X}}\wedge$ with $r$ derivatives like in
\begin{equation*}
\d^{(r)}(\co_{\hat{X}_1}\wedge)\cdots\,\d(\co_{\hat{X}_s}\wedge)\,,\qquad \d^{(r-1)}(\co_{\hat{X}_1}\wedge)\,\d'(\co_{\hat{X}_2}\wedge)\cdots\,\d(\co_{\hat{X}_s}\wedge)\,,\qquad \ldots\,\,.
\end{equation*}
In conclusion if $0<s<n$ the form degree $r$ can assume any integer value.

Finally, when the picture degree reaches the odd dimension $n$ of the supermanifold, the mechanism generating forms with negative degree $r$ is still valid because there is no obstruction in taking derivatives of $\d$-functions with $s=n\,$. On the contrary, now it is no longer possible to find any odd covector $\co_{\hat{X}}$ which is linearly independent from $s$ given independent covectors $\co_{\hat{X}_1},\ldots,\co_{\hat{X}_s}$. As a consequence the only exterior products one can use to increase the form degree are those with even arguments. The point is that, as these anticommute, one can apply at most $m$ of them. In conclusion if $s=n$ then $r$ must be lower or equal to the even dimension of the supermanifold, i.e. $r\le m$\,.

In order to study the structure of the various pictures, i.e. the subcomplexes $\Omega^{(\bullet|s)}$ with $0\le s\le n\,$, we kept the picture degree $s$ fixed and let the form degree $r$ vary within the intervals described above. In other words we have been moving horizontally in the supercomplex. Now we will introduce operators, called picture changing operators, which allow to move vertically as they modify the form degree leaving $s$ unchanged.

Given an odd scalar function $\hat{f}$ one can define the operator
\be\label{defY}
Y_{\hat{f}}\eqdef \hat{f}\,\,\delta(\de\hspace{-0.05mm}\hat{f}\wedge)\,.
\ee
It commutes with the differential,
\be
\left[\de,Y_{\hat{f}}\right]=0\,,
\ee
and it has degree $(0|1)$\,. Therefore it is a picture changing operator which maps cohomology classes of $\Omega^{(r|s)}$ to cohomology classes of $\Omega^{(r|s+1)}$\,. Furthermore the commutator of two picture changing operators $Y_{\hat{f}}$ and $Y_{\hat{g}}$ vanishes as easily follows from definition \eqref{defY}, hence the $Y$'s form a commutative algebra. Given an odd vector $\hat{X}$, one can define another operator
\be
Z_{\hat{X}}\eqdef\d(\intd_{\hat{X}})\,\lie_{\hat{X}}+\frac{1}{2}\,\d'(\intd_{\hat{X}})\,\intd_{[\hat{X},\hat{X}]}\,,
\ee
which, looking at \eqref{comddeltaintd} with $n=-1\,$, may also be expressed formally as
\be
Z_{\hat{X}}=\left[\de,\vartheta(\intd_{\hat{X}})\right],
\ee
where $\vartheta(x)$ is the Heavyside function satisfying $\vartheta'(x)=\d(x)\,$. It commutes with the differential like the $Y$'s,
\be
\left[\de,Z_{\hat{X}}\right]=0\,,
\ee
but it has degree $(0|-1)$\,. Therefore it is a picture changing operator which maps cohomology classes of $\Omega^{(r|s)}$ to cohomology classes of $\Omega^{(r|s-1)}$\,. Furthermore the commutator of two picture changing operators $Z_{\hat{X}}$ and $Z_{\hat{Y}}$ is
\be
\left[Z_{\hat{X}},Z_{\hat{Y}}\right]=\Big[\,\de\,,\,\delta(\intd_{\hat{X}})\,\delta(\intd_{\hat{Y}})\,\intd_{[\hat{X},\hat{Y}]}\,\Big]\,,
\ee
hence on the cohomology also the $Z$'s form a commutative algebra.

To be more explicit let us consider on a supermanifold of dimension $(m|n)$ the coordinate odd vectors $\,\vec{\partial}_\a$ ($\a=1,\ldots,n$) and the dual $1$-forms $\de\t^\a$\,. The corresponding picture changing operators will be
\begin{equation}\label{Yalpha}
\begin{gathered}
Y_\a=\t^\a\,\d(\de\t^\a\wedge)\,,\\
Z_\a=\d(\intd_\a)\,\de\,\intd_\a\,.
\end{gathered}
\end{equation}
For any $\a,\b=1,\ldots,n\,$, the operators $Y_\a$ and $Z_\a$ obey the commutation relations
\begin{equation}
\begin{gathered}
\left[Y_\a\,,\de\t^\b\wedge\,\right]=0\,,\\
\left[Z_\a\,,\intd_\b\right]=0\,,
\end{gathered}
\end{equation}
but in fact when $\a=\b$ one has
\begin{equation}\label{pcnull}
\begin{gathered}
Y_\a\,\,\de\t^\a\wedge=\de\t^\a\!\wedge\,Y_\a=0\,,\\
Z_\a\,\intd_\a=\intd_\a\,Z_\a=0\,.
\end{gathered}
\end{equation}
Consider now the picture changing operators of order $n$
\begin{equation}\label{Yalpha1n}
\begin{gathered}
Y_{1\,\ldots\,n}\eqdef\prod^{n}_{\a=1} Y_\a:\,\mathcal{H}^{(\cdot|0)}\longrightarrow\mathcal{H}^{(\cdot|n)}\,,\\
Z_{1\,\ldots\,n}\eqdef\prod^{n}_{\a=1} Z_\a:\,\mathcal{H}^{(\cdot|n)}\longrightarrow\mathcal{H}^{(\cdot|0)}\,.
\end{gathered}
\end{equation}
Because of \eqref{pcnull}, $Y_{1\,\ldots\,n}$ and $Z_{1\,\ldots\,n}$ will satisfy
\be\begin{gathered}
Y_{1\,\ldots\,n}\,\de\t^\a\wedge=\de\t^\a\!\wedge\,Y_{1\,\ldots\,n}=0\,,                     \\
Z_{1\,\ldots\,n}\,\intd_\a=\intd_\a\,Z_{1\,\ldots\,n}=0
\end{gathered}\ee
for any $\a=1,\ldots,n\,$. As a consequence $Y_{1\,\ldots\,n}$ annihilates all $(r|0)$-forms but the ordinary $r$-forms $\,\omega=\omega_{i_1\ldots i_r}(x)\,\de x^{i_1}\wedge\cdots\wedge\de x^{i_r}\,$ living on the body manifold, which are mapped to
\be\label{imgY}
Y_{1\,\ldots\,n}\,\omega=\omega_{i_1\ldots i_r}(x)\,\,\t^1\,\d(\de\t^1\wedge)\cdots\,\t^n\,\d(\de\t^n\wedge)\,\,\de x^{i_1}\wedge\cdots\wedge\de x^{i_r}\,.
\ee
On the other hand the inverse picture changing operator $Z_{1\,\ldots\,n}$ annihilates all elements of $\,\Omega^{(\bullet|n)}\,$ but those of the form $\,Y_{1\,\ldots\,n}\,\omega$\,, which are mapped back to the corresponding $r$-forms $\omega\,$. In conclusion the composite operators
\begin{equation}
\begin{gathered}
Z_{1\,\ldots\,n}\circ Y_{1\,\ldots\,n}\!:\,\Omega^{((r|0))}\longrightarrow\Omega^{((r|0))},\quad r\le m\,,\\
Y_{1\,\ldots\,n}\circ Z_{1\,\ldots\,n}\!:\,\Omega^{((r|n))}\longrightarrow\Omega^{(r|n))},\quad r\ge 0
\end{gathered}
\end{equation}
act as projectors on the $s=0$ and $s=n$ pictures respectively.

With different picture changing operators, for example $Y_\a=\t^\a\,\d(\de\t^\a\wedge)+\de x^j\wedge\,\d'(\de \t^\a\wedge)\,$, we would obtain other correspondences between cohomology classes. Nevertheless whatever choice one makes for $Y$ and $Z\,$, $Y$ cannot exist for negative form degree because $\Omega^{(r|0)}=0$ with $r<0$, and $Z$ cannot exist for form degree greater than the even dimension of the supermanifold because $\Omega^{(r|n)}=0$ with $r>m$. The structure of the supercomplex of forms is summarized in figure \ref{scomplex}\,.
\begin{figure}[t]
\begin{center}
\begin{tabular}{c@{\hskip -0mm}c@{\hskip 0mm}c}
\begin{tabular}{c@{\hskip 1.5mm}c@{\hskip 0.1mm}c}
&&\\[-0.34cm]
&\hspace{0.7cm}$0$&$\stackrel{\de}{\longrightarrow}$\\[0.06cm]
&\hspace{0.655cm}\tiny{$Z$}\hspace{0.3mm}\large{$\uparrow$}\phantom{\tiny{$Z$}}&\\[-0.05cm]
&\hspace{0.68cm}\vdots&\\[-0.05cm]
$\cdots$&$\Omega^{(-1|s)}$&$\stackrel{\de}{\longrightarrow}$\\[-0.15cm]
&\hspace{0.68cm}\vdots&\\[0.05cm]
&\hspace{0.655cm}\tiny{$Z$}\hspace{0.3mm}\large{$\uparrow$}\phantom{\tiny{$Z$}}&\\[0.05cm]
$\cdots$&$\Omega^{(-1|n)}$&$\stackrel{\de}{\longrightarrow}$\\[0.08cm]
\end{tabular}
&
\begin{tabular}{|@{\hskip 1.5mm}c@{\hskip 1mm}c@{\hskip 2mm}c@{\hskip 1.5mm}c@{\hskip 1.5mm}c@{\hskip 1.5mm}c@{\hskip 1.5mm}c@{\hskip 1.5mm}|}
\hline
&&&&&&\\[-0.34cm]
$\Omega^{(0|0)}$&$\stackrel{\de}{\longrightarrow}$&$\cdots$&$\Omega^{(r|0)}$ &$\cdots$&$\stackrel{\de}{\longrightarrow}$&$\Omega^{(m|0)}$\\[0.06cm]
\tiny{$Z$}\hspace{0.3mm}\large{$\uparrow$}\large{$\downarrow$}\hspace{0.5mm}\tiny{$Y$}&&&\tiny{$Z$}\hspace{0.3mm}\large{$\uparrow$}\large{$\downarrow$}\hspace{0.5mm}\tiny{$Y$}&&&\tiny{$Z$}\hspace{0.3mm}\large{$\uparrow$}\large{$\downarrow$}\hspace{0.5mm}\tiny{$Y$}\\[-0.05cm]
\vdots&&&$\vdots$&&&$\vdots$\\[-0.05cm]
$\Omega^{(0|s)}$&$\stackrel{\de}{\longrightarrow}$&$\cdots$&$\Omega^{(r|s)}$ &$\cdots$&$\stackrel{\de}{\longrightarrow}$&$\Omega^{(m|s)}$\\[-0.15cm]
\vdots&&&$\vdots$&&&$\vdots$\\[0.05cm]
\tiny{$Z$}\hspace{0.3mm}\large{$\uparrow$}\large{$\downarrow$}\hspace{0.5mm}\tiny{$Y$}&&&\tiny{$Z$}\hspace{0.3mm}\large{$\uparrow$}\large{$\downarrow$}\hspace{0.5mm}\tiny{$Y$}&&&\tiny{$Z$}\hspace{0.3mm}\large{$\uparrow$}\large{$\downarrow$}\hspace{0.5mm}\tiny{$Y$}\\[0.05cm]
$\Omega^{(0|n)}$&$\stackrel{\de}{\longrightarrow}$&$\cdots$&$\Omega^{(r|n)}$&$\cdots$ &$\stackrel{\de}{\longrightarrow}$&$\Omega^{(m|n)}$\\[0.08cm]
\hline
\end{tabular}
&
\begin{tabular}{c@{\hskip 2mm}c@{\hskip 1.5mm}c}
&&\\[-0.34cm]
$\stackrel{\de}{\longrightarrow}$&$\Omega^{(m+1|0)}$&$\cdots$\\[0.06cm]
&\hspace{-1.15cm}\phantom{\tiny{$Y$}}\hspace{0.5mm}\large{$\downarrow$}\hspace{0.5mm}\tiny{$Y$}&\\[-0.05cm]
&\hspace{-1.15cm}\vdots&\\[-0.05cm]
$\stackrel{\de}{\longrightarrow}$&$\Omega^{(m+1|s)}$&$\cdots$\\[-0.15cm]
&\hspace{-1.15cm}\vdots&\\[0.05cm]
&\hspace{-1.15cm}\phantom{\tiny{$Y$}}\hspace{0.5mm}\large{$\downarrow$}\hspace{0.5mm}\tiny{$Y$}&\\[0.05cm]
$\stackrel{\de}{\longrightarrow}$&\hspace{-1.15cm}$0$&\\[0.08cm]
\end{tabular}
\end{tabular}
\end{center}\vskip -0.2cm\caption{\rm Structure of the supercomplex of forms on a supermanifold of dimension $(m|n)\,$. The form degree $r$ changes going from left to right while the picture degree $s$ changes going from up to down. The rectangle contains the subset of the supercomplex where the various pictures are isomorphic.}\label{scomplex}
\end{figure}
The rectangle contains the region of $\Omega^{(\bullet|\bullet)}$ with $0<r<m$ where it is possible to define both $Y$ and $Z$\,.

The most general descending picture changing operator is a combination of the $Y_\a$ in \eqref{Yalpha} with operators of the form $\,\de x^{i_1}\wedge\cdots\wedge\de x^{i_{2h+1}}\wedge\de\t^{\a_1}\wedge\cdots\wedge\de\t^{\a_{k-2h-1}}\wedge\d^{(k)}(\de\t^\b\wedge)\,$, for example
\begin{gather*}
\de x^i\wedge\d'(\de\t^\a\wedge)\,,\quad \de x^i\wedge\de\t^a\wedge\d''(\de\t^\b\wedge)\,,\quad \de x^i\wedge\de x^j\wedge\de x^k\wedge\d'''(\de\t^\a\wedge)\,,\\
\de x^i\wedge\de\t^\a\wedge\de \t^\b\wedge\d'''(\de\t^\g\wedge)\,,\quad \de x^i\wedge\de x^j\wedge\de\t^\a\wedge\de x^k\wedge\d^{(4)}(\de\t^\b\wedge)\,.
\end{gather*}
Each $\de$-closed combination $Y$ of these terms realizes a cohomological map of the $s$-picture to the $(s+1)$-picture. By multiplying $n$ picture changing operators of order $1$ one obtains picture changing operators of order $n\,$ which are combinations of objects of the form
\begin{equation}\label{genpictcg}
\de x^{i_1}\wedge\ldots\wedge\de x^{i_k}\wedge\,(\t^1)^{\e_1}\cdots(\t^{n})^{\e_n}\,\d^{(h_1)}(\de\t^1\wedge)\cdots\,\d^{(h_n)}(\de\t^n\wedge)
\end{equation}
with $\,0\le k\le m\,$, $h_1+\cdots+h_n=k\,$, and $\e_\a=1-\text{sgn}(h_\a)\,$. Again each $\de$-closed combination $Y_{(n)}$ realizes a cohomological map between $(r|0)$-forms and $(r|n)$-forms. By including all possible terms like \eqref{genpictcg}, $Y_{(n)}$ will have a smaller null space than the picture changing operator $Y_{1\,\ldots\,n}$ given in \eqref{Yalpha1n}\,, but $Y_{(n)}$ will never be an isomorphism. Of course one can define many one to one correspondences between $\Omega^{(r|0)}$ and $\Omega^{(r|n)}\,$: an example is the map $W_{(n)}$ linearly generated by the action
\be
\begin{aligned}
W_{(n)}\big(\de x^{i_1}\wedge\ldots\wedge\de x^{i_k}\wedge(\de \t^1&\wedge)^{h_1}\ldots(\de \t^n\wedge)^{h_n}\big)\\&=
\de x^{i_1}\wedge\ldots\wedge\de x^{i_k}\wedge\,\d^{(h_1)}(\de\t^1\wedge)\cdots\,\d^{(h_n)}(\de\t^n\wedge)
\end{aligned}
\ee
where $k+h_1+\cdots\,h_n=m\,$. However this map neither preserves the form degree nor commutes with the differential, therefore $W_{(n)}$ does not map cohomology classes to cohomology classes. In general one can show that every picture is connected to each other by invertible maps but none of them preserve cohomology. A complete analysis of \v{C}eck and de Rham cohomology will be present elsewhere \cite{anto-marco}. 

Usually, one way to deal with delta functions is to use the dual space. Then, we can map
the superforms of $\Omega^{(n|r)}$ into the dual superforms by performing the Fourier
transform over the commuting $\de\psi_i$
\begin{equation}\label{ftA}
\hat\omega^{(n|r)}(\g,\psi_i, \de\g, \eta_i) = \int \omega^{(n|r)}(\g,\psi_i, \de\g, d\psi_i) e^{i \eta^i \de\psi_i}
\end{equation}
where $\eta_i$ are the dual variables to $\de\psi_i$. Since we have assumed that the superforms
are analytical functions of $\de\psi_i$ or delta distributions $\delta(\de\psi_i)$, it is easy to show that
the polynomial expressions $\omega^{(n|0)}$ are mapped into $\hat\omega^{(-n|2)}$ with
negative form number
\be\label{ftB}
\begin{aligned}
&\omega^{(n|0)} =  \Big(y_{p_1,p_2}(\g, \psi_i)+ f_{p_1,p_2}(\g, \psi_i) \, d\g \wedge\Big)
(\de\psi_1)^{p_1} \wedge (\de\psi_2)^{p_2} \longrightarrow\\
&\hat\omega^{(-n|2)}= \Big(y_{p_1,p_2}(\g, \psi_i)+ f_{p_1,p_2}(\g, \psi_i) \, d\g \wedge\Big)\delta^{(p_1)} (\eta_1)\wedge \delta^{(p_2)}(\eta_2)\,.
\end{aligned}
\ee
In the same way, we can map the superforms $\omega^{(n|2)}$ into $\hat\omega^{(-n|0)}$
(with negative $n$). This implies that the superforms with the delta functions are indeed needed also in the Fourier tranform picture.

\section{Integration}

Given an $m$-dimensional manifold $\,\mathcal{M}\,$ and a differential form $\,\varOmega\,$ of degree $m$\,, one can define the integral of $\,\varOmega\,$ over $\,\mathcal{M}\,$ by virtue of the fact that a top form transforms as a Jacobian and the result of the integration does not depend on the parametrization which is used. One could expect that the story can be extended to differential forms on supermanifolds simply by substituting usual forms with ``naive'' superforms. In fact it is easy to see that this is not the correct approach since top ``naive'' superforms do not even exist. The trick to give a meaningful notion of integral for differential forms over an $(m|n)$-dimensional supermanifold $\,\mathcal{\hat{M}}\,$ is to consider forms belonging to the $n$-picture instead of the $0$-picture. Indeed an $(m|n)$-form $\,\hat{\varOmega}\,$ is a top form transforming as a Berezinian and one can define the integral of $\hat{\varOmega}$ over $\mathcal{\hat{M}}$ as the Berezin integral
\begin{equation}\label{defsuperint}
\int_\mathcal{\hat{M}}\hat{\varOmega}\,\,\eqdef\,\int_\mathcal{\hat{M}}\mathcal{D}(x|\t)\,\,\,\hat{\varOmega}\!\left(X(x,\t)\,;\frac{\partial X}{\partial x^1}\,,\dots,\frac{\partial X}{\partial x^m}\,\right|\left.\frac{\partial X}{\partial \t^1}\,,\dots,\frac{\partial X}{\partial \t^n}\right).
\end{equation}
As in the case of ordinary forms the integral above does not depend on the parametrization $(x|\,\t)=(x^1,\ldots,x^m|\t^1,\ldots,\t^n)$ which is used on $\mathcal{M}\,$. Furthermore the Stokes theorem holds since, given an $(m-1|n)$-form $\xi$\,, the integral of $\de\xi$ over $\mathcal{M}$ is related to the integral of $\xi$ on the boundary $\partial\mathcal{M}$ by
\be
\int_{\mathcal{\hat{M}}}\de \xi=\int_{\partial\mathcal{\hat{M}}}\xi\,.
\ee
Given a top form $\,\varOmega=\varOmega(x)\,\,\de x^1\wedge\cdots\wedge\de x^m\,$ on the body manifold $\mathcal{M}$ and the top form $\,\hat{\varOmega}=Y_{1\,\ldots\,n}\,\varOmega=\varOmega(x)\,\,\t^1\,\d(\de\t^1\wedge)\cdots\,\t^n\,\d(\de\t^n\wedge)\,\,\de x^1\wedge\cdots\wedge\de x^m\,$ on the supermanifold $\hat{\mathcal{M}}\,$, from \eqref{defsuperint} one finds
\be\label{superident}
\int_{\hat{M}}\hat{\varOmega}=\int_\mathcal{M}\mathcal{D}(x|\t)\,\,\t^1\cdots\t^n\,\,\varOmega(x)\,=\int_{\mathcal{M}}\!\de^mx\,\,\varOmega(x)=\int_{\mathcal{M}}\varOmega\,.
\ee
This formula allows to write sistematically integrals over ordinary manifolds in terms of integrals over supermanifolds, and the operator $\,Y_{1\,\ldots\,n}\,$ is responsible for the localization showing once more the consintency of the adopted $\d$-function notation. This mechanism resembles the localization principle based on equivariant cohomology with the picture changing operator playing the role of a Thom form. In order to check whether the two constructions are really connected we now recall a few basic concepts about equivariant cohomology and the Thom class.
\subsection{Equivariant cohomology}
Let $\,\mathcal{M}\,$ be a $\,\mathcal{G}$-manifold, i.e. a manifold with an action $\,x\rightarrow g\,x\,$ for all $\,x\in\mathcal{M}\,$ and $\,g\in\mathcal{G}\,$. In this situation one seeks a notion of cohomology that would incorporate both the topology of the space and the action of the group. When the action of $\,\mathcal{G}\,$ is free, the quotient space $\,\mathcal{M}/\mathcal{G}\,$ forms the base space of a principal $\,\mathcal{G}$-bundle
\be
\begin{CD}
\M@<<<\G\\
@VV\pi V\\
\M/\G
\end{CD}
\ee
and one can study the cohomology of $\,\mathcal{M}/\mathcal{G}\,$ in the usual sense. However in general the group action has fixed points and the quotient space is not even a manifold. In this case the standard trick is to replace $\M$ with $\,\E\G\times\M\,$, where $\E\G\rightarrow\B\G\,$ is the universal $\G$-bundle. Since $\E\G$ is contractible, $\,\E\G\times\M\,$ is homotopically equivalent to $\M$ and the action of $\G$ on the product space is free because it is on $\,\E\G\,$. Therefore
\be
\E\G\times_{\G}\M\,\eqdef\,\frac{\E\G\times\M}{\G}
\ee
can be regarded as a bundle over $\B\G$ with fiber $\M$ and one can study the cohomology of $\E\G\times_{\G}\M\,$ to incorporate the topology and the group action. This leads to the definition of the topological $\G$-equivariant cohomology of $\M$\,:
\be
\mathcal{H}_{\,\G,\text{top}}^{\,\bullet}(\M)\eqdef\mathcal{H}^{\,\bullet}\!\left(\E\G\times_{\G}\M\right).
\ee

Although $\E\G$ is homotopically trivial, it is nontrivial as a bundle over $\B\G$, and the elements of the cohomology $\H^{\,\bullet}(\B\G)$ called characteristic classes measure the twisting of the bundle. Furthermore one can measure the topology of any $\G$-bundle $\P\rightarrow\M$ by pullback from $\H^{\,\bullet}(\B\G)$ since the fundamental property of $\,\E\G\rightarrow\B\G\,$ is that one can find a copy of any $\G$-bundle sitting inside it:
\be
\begin{CD}
\hspace{-0.7cm}\P=f^*\,\E\G &&\E\G\\
@VVV @VVV\\
\M @>f>>\B\G
\end{CD}\,\,\,.
\ee
Given a connection $A\in\varOmega^1(\P,\mfg)$\, on $\P$\,, characteristic classes $\widetilde{ch}_k$ can be obtained from the field strengths $F=\de A+\half\,[A,A]\in\varOmega^2(\P,\mfg)$\, through the formula
\be
\widetilde{ch}_k=\frac{\text{Tr}\,{F^k}}{(2\pi i)^k\,k!\,}\in \varOmega^{2n}(\P)\,.
\ee
But it is easy to see that the $\widetilde{ch}_k$'s are closed and therefore exact since $\E\G$ is contractible. The $\widetilde{ch}_k$'s neither have vertical component nor vary under the action of a vertical vector field $\xi(X)$ with $X\in\mfg$\,,
\be\label{character}
\intd_{\,\xi(X)}\,\widetilde{ch}_k=0\,,\qquad\lie_{\,\xi(X)}\,\widetilde{ch}_k=0\,.
\ee
Forms satisfying these conditions are said to be basic because they can always be obtained by pullback from forms living on the base space. In particular there must be some forms $ch_k$ such that $\widetilde{ch}_k=\pi^*(ch_k)\,$. The new forms $ch_k$ define nontrivial cohomology classes on $\B\G$ which do not depend on the connection $A$\,.

To get the corresponding algebraic description of equivariant cohomology one has to introduce the Weil algebra $\W(\mathfrak{g})$ of $\mathfrak{g}\,$, that is the differential graded algebra
\be
\W(\mfg)=S(\mfg^*)\otimes\L(\mfg^*)
\ee
where $S(\cdot)$ and $\L(\cdot)$ denote the symmetric algebra and the exterior algebra respectively. $\W(\mfg)$ may be described in terms of a set of generators $\{\phi^i\}$ of degree $2$ for $S(\mfg^*)$ and a set of generators $\{\psi^i\}$ of degree $1$ for $\L(\mfg^*)$ together with the differential $\de_{\W}$ defined by
\begin{equation}
\begin{gathered}
\de_{\W}\,\psi^i=\phi^i-\half\,\,f^i_{~jk}\,\psi^j\psi^k,\\
\de_{\W}\,\phi^i=-\,f^i_{~jk}\,\psi^j\phi^k
\end{gathered}
\end{equation}
where $f^i_{~jk}\,$ are the structure constants of $\mfg$\,. $\W(\mfg)$ is the algebraic analog of $\E\G$. Just as $\E\G$ is contractible, the cohomology of the Weil algebra is trivial:
\be
\H^{\,\bullet}(\W(\mfg),\de_{\W})=\d_{\bullet,0}\,\mathbb{R}\,.
\ee
To get interesting cohomology one can follow the same path as for characteristic classes and introduce on $\W(\mfg)$ an interior product $I_i$ and a Lie derivative $L_i$ defined by the following action on the generators:
\begin{equation}
\begin{gathered}
I_i\,\psi^j=\d_{~i}^{\,j}\,,\\
I_i\,\phi^j=0\,,\\
L_i\eqdef\{I_i,\de_{\W}\}\,.
\end{gathered}
\end{equation}
An element of $\W(\mfg)$ will be called basic if it is annihilated both by the $I_i$'s and the $L_i$'s similarly to \eqref{character}. The basic subcomplex $\B\mfg$ of $\W(\mfg)$ consists of the invariant polynomials on the Lie algebra $\mfg$\,, i.e. $\B\mfg$\, is the algebra of Casimir invariants. The analog of the replacement $\M\rightarrow\E\G\times\M\,$ of topological equivariant cohomology is the substitution $\varOmega(\M)\rightarrow\W(\mathfrak{g})\otimes\varOmega(\M)\,$ and the algebraic $\G$-equivariant cohomology of $\M$ is
\be
\H^{\,\bullet}_{\,\G,\text{alg}}(\M)\eqdef\H^{\,\bullet}((\W(\mfg)\otimes\varOmega(\M))_{\text{basic}}\,,\de_{\text{tot}})
\ee
where the total differential is
\be\label{totdiff}
\de_{\text{tot}}=\de_{\W}\otimes 1+1\otimes\de
\ee
and the basic subcomplex contain forms $\omega$ which satisfy the horizontality and invariance conditions
\be\label{basicsubc}
(I_i\otimes 1+1\otimes\intd_{\xi(X_i)})\,\omega=0\,,\qquad(L_i\otimes 1+1\otimes\lie_{\xi(X_i)})\omega=0\,.
\ee
The relationship between algebraic and topological equivariant cohomology is much more than a similarity, indeed for $\G$ compact one has
\be
\H^{\,\bullet}_{\,\G,\text{alg}}(\M)=\mathcal{H}_{\,\G,\text{top}}^{\,\bullet}(\M)\,.
\ee

\subsection{Bosonic Thom class}
Given an orientable vector bundle $\,\E\rightarrow\M\,$ of rank $\,n$\, with fiber $V$ endowed with a nondegenerate pairing $(\,\cdot\,,\,\cdot\,)_V\,$, integration along the fibers defines an isomorphism between the cohomology on the base space and the cohomology with rapid decrease along the fibers:
\be\label{Thomiso}
\pi_*:\,\H^{\,\bullet}_{\scriptscriptstyle VRD}(\E)\cong\H^{\,\bullet\,-n}(\M)\,.
\ee
The image of $1\in\H^{\,0}(\M)$ under $\pi_*^{-1}$ defines a cohomology class called Thom class,
\be
\pi^{-1}_*(1)\eqdef\Phi(\E)\in \H^{\,n}_{\scriptscriptstyle VRD}(\E)\,,
\ee
which allows to rewrite the Thom isomorphism \eqref{Thomiso} as
\be
\begin{gathered}
\mathcal{T}:\,\H^{\,\bullet}(\M)\rightarrow \H^{\,\bullet+n}_{\scriptscriptstyle VRD}(\E)\\
\omega\mapsto\pi^*(\omega)\wedge\Phi(\E)\,.
\end{gathered}
\ee
One of the nice properties of the Thom class is the following localization principle. If $s$ is a generic section of $E$ and $\mathcal{Z}(s)\stackrel{_i}{\hookrightarrow}\M$ its zero locus then
\be\label{localization}
\int_{\mathcal{Z}(s)}i^*\omega=\int_{\M}s^*(\Phi(\E))\wedge\omega\,.
\ee
Therefore it is important to construct representatives for $\Phi(\E)$. The standard technique to do it is to realize a universal representative by replacing constructions on the twisted bundle $\E$ with equivariant constructions on a trivial $\SO(V)$-bundle, and then to get elements of the Thom class in an explicit form by mapping the universal representative back to $\mathcal{E}$. It turns out that a universal Thom form $U$ is an element of the basic subcomplex of $\W(\text{so}(V))\otimes\varOmega(V)$ which is closed with respect to the total differential in \eqref{totdiff} and such that the integral along the fibers is $\int_VU=1\,$.

Integral representatives of the universal Thom class can be obtained by introducing for $V^*$ a set of orthonormal coordinates $\rho_a\,$, which anticommute as the $\de x$'s. They are interpreted as ``antighosts'' in a field theoretic context. Consider the form $U\in\W(\text{so}(V))\otimes\varOmega(V)$ given by
\be\label{Thomformb}
U=\frac{1}{\pi^n}\,\,e^{-(x,x)_V}\!\int_{V^*}\de\r \,\,\,e^{\,\frac{1}{4}(\r,\phi\r)_{V^*}+\,i\,\langle\nabla x,\,\r\rangle}
\ee
where $x^a$ are the coordinates on $V$, $\langle\,\cdot\,,\,\cdot\,\rangle$ is the canonical pairing, $\nabla x$ denotes $\de x+A\cdot x$ with $A$ a connection on the fiber and $\phi$ is the curvature $\de A+A\wedge A$\,. It is a Thom form. Indeed it is horizontal by construction and, as it is easy to see, its integral on $V$ is 1\,:
\be
\begin{aligned}
\int_VU\,&=\frac{1}{\pi^n}\,\int_V\int_{V^*}\de\r\,\,\,e^{-(x,x)}\,\frac{i^{\,2n}}{(2n)!}\,\,(\de x^a\r_a)^{2n}\\
&=\frac{1}{\pi^n}\,\int_V\de x^1\wedge\cdots\wedge\de x^{2n}\,\,e^{-(x,x)}\,=1\,.
\end{aligned}
\ee
One can also prove that $U$ is closed by rewriting it in a BRST integral representation where we have a BRST operator $Q$ acting according to
\begin{equation}\label{BRStransf}
\begin{gathered}
QA=\phi\,, \quad Q\phi=0\,,\quad Qx=\nabla x\,,\quad
Q\rho=\pi\,, \quad Q\pi=\phi\rho\,.
\end{gathered}
\end{equation}
Notice that we adopt the usual identification $dx \equiv {\rm ghost}$. In this way the 
ghost number is the form degree. Using this convention, we get that $A$ has ghost number 
1 and $\phi$ has number 2. In addition, it is easy to see that the BRST is nilpotent using 
the fact that $\nabla^2 = \phi$. 
In this representation $U$ takes the form
\begin{equation}\label{BRSrep}
U=\int_{V^*\times\Pi
V^*}\prod_{\alpha=1}^{2n}\frac{\de\pi_\alpha}{\sqrt{2\pi}}\,\frac{\de\rho_\alpha}{\sqrt{2\pi}}\,\,e^{\,Q\Psi}
\end{equation}
where $\Psi$ is the gauge fermion given by
\begin{equation}\label{gaugeferm}
\Psi=-i\,\langle\rho,x\rangle-(\rho,\pi)_{\scriptscriptstyle V^*}\,.
\end{equation}
A complete analysis has been presented in the pedagogical account \cite{Cordes:1994fc} and 
refer to it for details. 

%%%%%%%%%%%%%%%%%%%%%%%%%%%%%%%%%%%%%%%%%%%%%%%%%

\subsection{Fermionic Thom class}
We now consider a fermionic counterpart of the ordinary Thom class illustrated above to give a prescription for the integration on supermanifolds. In our construction the vector bundle $\E$ introduced in last section will be replaced by a Riemannian supermanifold $\hat{\M}$ of odd dimension $2n$ and the fiber $V$ by the soul space $\hat V$ of $\hat{\M}$\,. Actually $\hat{\M}$ is not just a vector bundle but a ringed superspace endowed with a sheaf structure generated by a Grassmann algebra, therefore it admits a wider class of morphisms than a simple vector bundle where one has just linear transformations connecting one fiber to another. Nevertheless the construction of the Thom class follows basically the same steps as in the bosonic case, except for the fact that the coordinates $\t^\m$ on $\hat V$ are anticommuting and the natural pairing $(\cdot\,,\cdot)_{\mathcal{S}}$ given by the fermionic supermetric $G_{\m\n}$ of $\hat{\M}$ is now antisymmetric. For that reason we denote by $\hat V$ the anticommuting "fiber" whose coordinates are $\t$'s.

On a supermanifold $\hat{\M}^{(m|2n)}$ the Thom class $\Phi(\hat{\M})$ relates integrals on $\hat{\M}$ to ordinary integrals on the body manifold $\M$ according to
\begin{equation}
\int_{\M}i^*\omega=\int_{\hat{\M}}s^*\!\left(\Phi(\hat{\M})\right)\wedge\omega
\end{equation}
where $\,i\,$ is the embedding of $\M$ into $\hat{\M}$\,, $\M\stackrel{_i}{\hookrightarrow}\hat{\M}$\,, and $s$ is the soul projection (see \cite{vara} and \cite{catenacci} for its 
definition), whose zero locus is the body $\M$\,, $\mathcal{Z}(s)=\M\,$. Therefore the Thom class $\Phi(\mathcal{M})$ acts as a picture changing operator of order $2n$ and defines an isomorphism $\mathcal{T}$ between the cohomology $\H^{\,\bullet}(\M)$ and the supercohomology $\H^{(\bullet|2n)}(\hat{\M})$\,:
\begin{equation}
\begin{gathered}
\mathcal{T}:\H^\bullet(\M)\longrightarrow\H^{(\bullet|2n)}(\hat{\M})\\
\omega\mapsto\pi^*(\omega)\wedge\Phi(\hat{\M})
\end{gathered}
\end{equation} 
with $\pi$ denoting the body projection. In other words it is possible to identify the cohomology on the body manifold $\M$ with the cohomology on the supermanifold $\hat{\M}$ via the integration $\pi_*$ on the soul space:
\begin{equation}
\pi_*:\H^{\bullet}(\M)\cong\H^{(\bullet|2n)}(\hat{\M})\,.
\end{equation}
In particular the Thom class will be the image of $1\in\H^0(\M)$ under $\pi_*^{-1}$\,:
\begin{equation}
\pi_*^{-1}(1)=\Phi(\hat{\M})\,.
\end{equation}

In order to write representatives of $\Phi(\hat{\M})$ for practical calculations we start from the same BRST integral representation of \eqref{BRSrep}, with a gauge fermion formally identical to \eqref{gaugeferm}\,,
\begin{equation}
\Psi=-i\,\langle\rho,\t\rangle-(\rho,\pi)_{\scriptscriptstyle V^*}\,,
\end{equation}
and with the BRST transformations
\begin{equation}
Q\omega^A=\phi^A\,,\quad Q\phi^A=0\,,\quad Q\t^\a=\nabla\t^\a\,,\quad
Q\rho_\a=\pi_\a\,,\quad Q\pi_\a=(\phi\rho)_\a
\end{equation}
where $\nabla \t$ denotes $\nabla\t=\de\t+\omega\cdot\t$ with $\omega$ the Riemannian superconnection and $\phi$ is the super-Riemann tensor. 
These transformations are the same as \eqref{BRStransf} except for the fact that the parity of the momenta $\pi$ and of the antighosts $\rho$ are opposite with respect to the bosonic case since the $\t$'s anticommute. So the $\pi$'s are odd while the $\rho$'s are even. From now on we will denote the covariant differentials $\nabla\t^\a$ by $\lambda^\a$ as usual in topological strings and supersymmetry \cite{Hori:2003ic}. Again we used the conventional identification of differential form and ghosts, therefore 
the BRST charge has ghost number one as it should be. 
We have to notice that the $\l$'s are complex commuting quantities and, as is clear from the previous section, we will consider analytic expressions in terms of those.\footnote{One can even conceive algebraic curves in the space of covariant differentials $\nabla\t$\,. Pure spinors of \cite{Berkovits:2000fe} are examples of costrained differentials.} In general the Thom form one finds with this procedure is given by
\begin{equation}\label{Thom-int-comp}
\begin{aligned}
U&=\int_{\hat V^*\times\Pi
\hat V^*}\prod_{\alpha=1}^{2n}\frac{\de\pi_\alpha}{\sqrt{2\pi}}\,\frac{\de\rho_\alpha}{\sqrt{2\pi}}\,\,e^{\,Q\Psi}\\
&=\frac{1}{(2\pi)^{2n}}\int\de\pi\,\de\rho\,\,\,
e^{\,-i\langle\pi,\,\theta\rangle-i\langle\rho\,,\lambda\rangle-(\pi,\pi)_{\hat V^*}-(\rho,\phi\rho)_{\hat V^*}}\\
&=\frac{1}{\left(2\pi\right)^n}\frac{\text{Pfaff}\!\left(g^{-1}\right)}{\sqrt{\det\!\left(g^{-1}\phi\right)}}\,\,\,e^{\,-\frac{1}{4}(\t,\t)_{\hat V}+\frac{1}{4}(\lambda\phi^{-1}\!,\,\lambda)_{\hat V}}\,,
\end{aligned}
\end{equation}
which, on complex supermanifolds, takes the form
\begin{equation}
U=\frac{1}{\left(2\pi\right)^n}\frac{1}{\det(\phi)}\,\,\,e^{\,-\frac{1}{2}(\t,\bar{\t})_{\hat V}+\frac{1}{2}(\lambda\phi^{-1}\!,\,\bar{\lambda})_{\hat V}}\,.
\end{equation}
Integral representatives of the universal Thom class can also be obtained by considering the form $U\in\W(\text{sp}(\hat V))\otimes\varOmega(\hat V)$ given by
\be\label{intrho}
U=\frac{1}{2^n\pi^{2n}}\,\,e^{\frac{1}{4}(\t,\t)_V}\!\int_{V^*}\de\r \,\,\,e^{\,-(\r,\phi\r)_{V^*}-\,i\,\langle\r,\nabla \t\rangle}\,,
\ee
which is the fermionic version of \eqref{Thomformb}.
 It is easy to see that it fulfills the horizontality and normalization properties of a Thom form. Indeed its integral on $\hat V$ is 1\,:
\be
\int_VU\,=\frac{1}{2^n\pi^{2n}}\,\int_V\de\t\int_{V^*}\de\r\,\,\frac{1}{4^nn!}\,(\t,\t)^n\,(2\pi)^{2n}\,\delta(\de\t^1\wedge)\cdots\delta(\de\t^{2n}\wedge)=1\,.
\ee

When the fermionic curvature  has flat directions one cannot write the Thom form as in \eqref{Thom-int-comp} as $\phi$ is not invertible. This is the case of $\mathbb{CP}^{(1|2)}$, whose supercurvature is given by
\begin{equation}\label{ignor}
\begin{aligned}
\phi^{\,~\nu}_{\mu}&=R^{\,~\nu}_{\mu~\rho\bar{\sigma}}\,\lambda^\rho\lambda^{\bar{\sigma}}\\
&=-\frac{1}{1+z\bar{z}+\t\bar{\t}}\,\left(\,(\lambda\bar{\lambda})\,\delta^{~\nu}_{\mu}-\lambda^{\bar{\m}}\lambda^\nu-\frac{(\lambda\bar{\t})(\t\bar{\l})}{1+z\bar{z}+\t\bar{\t}}\,\,\delta^{~\nu}_{\mu}+\frac{(\t\bar{\l})}{1+z\bar{z}+\t\bar{\t}}\,\,\t^{\bar{\mu}}\lambda^\nu
\right)
\end{aligned}
\end{equation}
and it has one flat direction. In the following calculation we need the inverse of the fiber metric,
\begin{equation}
g^{\bar{\nu}\rho}=(1+z\bar{z}+\t\bar{\t})\left(\delta^{\,\bar{\nu}\rho}+\frac{\t^{\bar{\n}}\t^\rho}{1+z\bar{z}}\right)\,.
\end{equation}
To define the integral of \eqref{intrho} we split the integration into the flat direction and a second independent one. $\rho$ may be written as a sum of two terms one of which is along the flat direction
\begin{equation}\label{decomp}
\rho_\a=\rho_\|\,g_{\a\bar{\b}}\l^{\bar{\b}}+\rho_\perp\epsilon_{\a\b}\l^\b
\end{equation}
By substituting \eqref{decomp} into the integral \eqref{Thom-int-comp} it yields
\begin{equation}\label{mappaz}
\begin{aligned}
U=&\,\,\frac{e^{-\frac{1}{2}(\t,\bar{\t})_V}}{4\pi^4\,\det(g)}\,\int\de\rho\,\de\bar{\rho}\,\,\,e^{
-i\langle\rho,\lambda\rangle
-i\langle\bar{\rho},\bar{\lambda}\rangle-2(\bar{\rho},\phi\rho)_{V^*}}\\
=&\,\,\frac{1+z\bar{z}}{4\pi^4}\,\,\left(\lambda\bar{\lambda}-\frac{(\lambda\bar{\t})(\t\bar{\lambda})}{1+z\bar{z}+\t\bar{\t}}\right)\,\,e^{-\frac{\t\bar{\t}}{2(1+z\bar{z})}\left(1-\frac{2\t\bar{\t}}{1+z\bar{z}}\right)}\\&\,\,\cdot\int\de\rho_\|\,\de\rho^*_\|\,\,e^{\,i(\rho_\|+\rho_\|^*)\left(\frac{\lambda\bar{\lambda}}{1+z\bar{z}+\t\bar{\t}}-\frac{(\lambda\bar{\theta})(\theta\bar{\lambda})}{(1+z\bar{z}+\t\bar{\t})^2}\right)}\\
&\,\cdot\int\de\rho_\perp\de\rho^*_\perp\,\,e^{-2\left|\rho_\perp\right|^2(\lambda\bar{\lambda})^2\left(\lambda\bar{\lambda}-\frac{(\lambda\bar{\t})(\t\bar{\lambda})}{1+z\bar{z}+\t\bar{\t}}-\frac{(\lambda\bar{\lambda})(\theta\bar{\theta})^2}{2(1+z\bar{z})^2}+\frac{(\bar{\lambda}\cdot\bar{\theta})(\theta\cdot\lambda)}{1+z\bar{z}}\right)}\\
=&\,\,\frac{1+z\bar{z}}{4\pi^3(\lambda\bar{\lambda})}\,\,\frac{\lambda\bar{\lambda}-\frac{(\lambda\bar{\t})(\t\bar{\lambda})}{1+z\bar{z}+\t\bar{\t}}}{\lambda\bar{\lambda}-\frac{(\lambda\bar{\t})(\t\bar{\lambda})}{1+z\bar{z}+\t\bar{\t}}-\frac{(\lambda\bar{\lambda})(\theta\bar{\theta})^2}{2(1+z\bar{z})^2}+\frac{(\bar{\lambda}\cdot\bar{\theta})(\theta\cdot\lambda)}{1+z\bar{z}}}\,\,e^{-\frac{\t\bar{\t}}{2(1+z\bar{z})}\left(1-\frac{2\t\bar{\t}}{1+z\bar{z}}\right)}\\&\,\,\cdot\int\de\rho_\|\,\de\rho^*_\|\,\,e^{\,i(\rho_\|+\rho_\|^*)\left(\frac{\lambda\bar{\lambda}}{1+z\bar{z}+\t\bar{\t}}-\frac{(\lambda\bar{\theta})(\theta\bar{\lambda})}{(1+z\bar{z}+\t\bar{\t})^2}\right)}
\end{aligned}
\end{equation}
There is a pole in $\lambda\bar{\lambda}$. This is expected in contrapposition to the bosonic case. In that 
case due to the Berezin integral the Thom class is a polynomial in the usual forms. In the case of supermanifolds, since we are admitting analytical expression in the superforms $\de\theta^\a$, it 
follows that also poles are admitted. In the next section, we provide a regularization technique 
by adding a new BRST multiplet and a gauge fermion. 

\subsection{Regularization}
The integral in \eqref{mappaz} cannot be computed directly due to the presence of a flat direction, i.e. a gauge symmetry. Therefore we use the conventional BRST technology to select a given gauge slice where we perform the integration. The result turns out to be independent of this choice if BRST invariant quantities are evaluated \cite{Cordes:1994fc}. We rewrite the integral as follows 
\begin{equation}\label{intreg}
\int\de\rho_{\alpha}\,\de\rho_{\bar{\alpha}}\,\,e^{-i(\rho_\a\lambda^\a+\rho_{\bar{\a}}\lambda^{\bar{\a}})-\frac{1}{2}\,\rho_{\bar{\a}}\,g^{\bar{\a}\b}\phi_\b^{\,\,\,\gamma}\rho_\gamma}
\end{equation}
where the gauge symmetry
\begin{equation}
\delta\rho_\a=\eta \,g_{\a\bar{\b}}\lambda^{\bar{\b}}\,,\qquad\qquad\delta\rho_{\bar{\a}}=-\,\bar{\eta} \,\lambda^{\b}g_{\b\bar{\a}}\,,
\end{equation}
is manifest. In fact, using
\begin{equation}
\phi_\b^{\,\,\,\gamma}(\delta\rho_\gamma)=\phi_\b^{\,\,\,\gamma}\left(\eta\,g_{\gamma\bar{\b}}\lambda^{\bar{\b}}\right)=0\,,\qquad\qquad \delta\rho_{\bar{\gamma}}\,g^{\bar{\gamma}\b}\phi_\b^{\,\,\,\gamma}=0
\end{equation}
which follows from (\ref{ignor}), 
we get
\begin{equation}
-i(\eta\,g_{\a\bar{\b}}\lambda^{\bar{\b}}\lambda^\a-\bar{\eta}\,\lambda^\b g_{\b\bar{\a}}\lambda^{\bar{\a}})=-i(\eta-\bar{\eta})\lambda^\a g_{\a\bar{\b}}\lambda^{\bar{\b}}=0
\end{equation}
iff $\bar{\eta}=\eta$. Therefore the action is invariant if the gauge parameter is real.
By promoting $\eta$ to a ghost c we have the BRST symmetry
\be
s\rho_\a=c\,g_{\a\bar{\b}}\lambda^{\bar{\b}}\,,\qquad s\rho_{\bar{\a}}=-c\,\lambda^\b g_{\b\bar{\a}}\,,\qquad sc=0\,.
\ee
As usual we add an auxiliary doublet with BRST transformations $sb=\bar{c}$\,, $s\bar{c}=0$\,. Hence the integral \eqref{intreg} is replaced by
\begin{equation}
\int\de c\,\de\bar{c}\,\de b \int\de\rho_{\alpha}\,\de\rho_{\bar{\alpha}}\,\,e^{-i(\rho_\a\lambda^\a+\rho_{\bar{\a}}\lambda^{\bar{\a}})-\frac{1}{2}\,\rho_{\bar{\a}}\,g^{\bar{\a}\b}\phi_\b^{\,\,\,\gamma}\rho_\gamma+s[\Psi]}
\end{equation}
where we have chosen the gauge fermion
\begin{equation}
\Psi=\left|\rho\right|^2\bar{c}\,\left(1+\frac{\a b}{2}\right)\,.
\end{equation}
Its variation is easily computed
\begin{equation}
\begin{aligned}\label{sPsi}
s\Psi&=c\left(g_{\a\bar{\b}}\lambda^{\bar{\b}}(g^{\a\bar{\gamma}}\rho_{\bar{\gamma}})-\rho_\a g^{\a\bar{\b}}\lambda^{\b}g_{\b\bar{\b}}\right)\bar{c}\,\left(1+\frac{\a b}{2}\right)+\left|\rho\right|^2\bar{c}\,\left(b+\frac{\a b^2}{2}\right)\\
&=-\left(\rho_\a\lambda^\a-\rho_{\bar{\a}}\lambda^{\bar{\a}}\right)c\bar{c}\left(1+\frac{\a b}{2}\right)+\left|\rho\right|^2\!\left(b+\frac{\a b^2}{2}\right)\,.
\end{aligned}
\end{equation}
Finally we compute the integral over b
\begin{equation}
-\left(\rho_\a\lambda^\a-\rho_{\bar{\a}}\lambda^{\bar{\a}}\right)c\bar{c}\,\frac{\a}{2}+\left|\rho\right|^2\!\left(1+\a b\right)=0
\end{equation}
which gives
\begin{equation}
b=-\frac{1}{\a|\rho|^2}\,\left(|\rho|^2-\frac{\a}{2}\left(\rho_\a\lambda^\a-\rho_{\bar{\a}}\lambda^{\bar{\a}}\right)\!c\bar{c}\right)
\end{equation}
and by plugging it into \eqref{sPsi} it yields
\begin{equation}
s\Psi=\frac{1}{2}\left(\rho_\a\lambda^\a-\rho_{\bar{\a}}\lambda^{\bar{\a}}\right)\!c\bar{c}-\frac{1}{2\a}\left|\rho\right|^2
\end{equation}
and the compete integral is 
\begin{equation}\label{polpetta}
\int\de c\int\de\bar{c}\int\de\rho_\a\de\rho_{\bar{\a}}\,\,e^{-i\left(1+\frac{i}{2}c\bar{c}\right)\rho_\a\lambda^\a-i\left(1-\frac{i}{2}c\bar{c}\right)\rho_{\bar{\a}}\lambda^{\bar{\a}}-\frac{1}{2}\rho_{\bar{\a}}g^{\bar{a}\b}\left(\phi_\b^{\,\,\,\gamma}+\frac{1}{\a}\delta_\b^{\,\,\gamma}\right)\rho_\b}\,.
\end{equation}
Note that the term $\,\phi_\b^{\,\,\,\gamma}+\frac{1}{\a}\delta_\b^{\,\,\gamma}\,$ is the gauge-fixed two point function. This makes the integral gaussian avoiding the pole in $\lambda\bar{\lambda}$. 
We conlcude with some remarks: we succeded to obtain a meaningful expression for the 
fermionic Thom class for ${\mathbb CP}^{(1|2)}$. The same technology can be adopted for any 
manifold embedded into a supermanifold. The resulting expression (\ref{polpetta}) has no singularity 
and it can be used to path integral computations. 

\section{Embedding of a K\"ahler manifold into a SCY}

A Calabi-Yau 3-fold $\mathcal{M}$ is completely characterized by a
pair of objects: the K\"ahler $2$-form $K$ and the holomorphic
$3$-form $\varOmega$. One can easily embed $\mathcal{M}$ into a CY
supermanifold $\mathcal{\hat{M}}$ of dimension $(3|4)$ by defining
a K\"ahler superform
\begin{subequations}\label{embed1}
\be
\label{cyA} \hat K = K_{m \bar n}\, \de z^{m} \wedge \de z^{\bar
n} + \,\theta^{\m} \theta^{\bar \n}\,
\delta(d\theta^{\m}\wedge)\,\delta(d\theta^{\bar \n}\wedge) \ee
and a holomorphic top form
\be
\label{cyB}
\begin{aligned}
\hat{\varOmega} = \varOmega_{\,mnp}\, \theta^{\m}  \theta^{\n}
\theta^{\rho} \theta^{\sigma} \,\de z^{m} \wedge& \de z^{n} \wedge
\de z^{p} \wedge \\& \delta(\de\theta^{\m}\wedge)\,
\delta(\de\theta^{\n}\wedge)\, \delta(\de\theta^{\r}\wedge)\,
\delta(\de\theta^{\s}\wedge)\,.
\end{aligned}
\ee
\end{subequations}
It can be checked that they are both closed and $\hat K\wedge \hat
\varOmega =0\,$. We can now integrate $\hat K$ and $\hat
\varOmega$ on the CY supermanifold $\hat{\cal M}$\,. Remembering
the integration rule \eqref{superident} it is easy to see that the
results coincide with the original bosonic integrals:
\begin{subequations}\label{cyD}
\begin{gather}
\begin{aligned}
\frac{1}{35\cdot(4!)^2}\,\int_{{\cal \hat M}} \underbrace{\,\hat
K\wedge\dots \wedge \hat K\,}_{7} &= \int_{{\cal \hat M}} K\wedge
K \wedge K\wedge \hat U\\&=\int_{{\cal M}} K\wedge K \wedge K\,,
\end{aligned}
\\[0.2cm]
\int_{{\cal \hat M}} \hat \varOmega \wedge \overline{\hat
\varOmega} = \int_{\cal \hat M} \varOmega \wedge
\overline{\varOmega}\wedge \hat U=\int_{\cal M} \varOmega \wedge
\overline{\varOmega}\,.
\end{gather}
\end{subequations}
Notice that in both cases the dimension of the forms in the
integrals coincides with the dimension $(3|4)$ of the SCY. Indeed
in the second example one needs seven powers of $\hat K$ to reach
the total dimension of $\mathcal{\hat M}\,$. In conclusion it is
possible to embed a CY manifold into a SCY with flat fermionic
directions without changing the results. Let us try to modify the
present situation by considering K\"ahler manifolds which are not
Ricci-flat. We know from \eqref{Ksuperext} how to embed them into
Calabi-Yau supermanifolds. Instead of \eqref{embed1}, we now set
\begin{subequations}\label{embed2}
\be
\hat K = K_{m \bar n}(z,\bar{z},\t,\bar{\t})\, \de z^{m}\!\wedge
\de z^{\bar n} + \,H_{\m\bar \n}(z,\bar{z},\t,\bar{\t})\,
\delta(d\theta^{\m}\wedge)\,\delta(d\theta^{\bar \n}\wedge) \ee
for the K\"ahler form and
\be
\begin{aligned}
\hat{\varOmega} = \varOmega_{\,mnp\,\m\n\r\s} \,\,\de z^{m}
\wedge& \de z^{n} \wedge \de z^{p} \wedge \\&
\delta(\de\theta^{\m}\wedge)\, \delta(\de\theta^{\n}\wedge)\,
\delta(\de\theta^{\r}\wedge)\, \delta(\de\theta^{\s}\wedge)\,.
\end{aligned}
\ee
\end{subequations}
for the $3$-form, which corresponds to consider curved superspace.
Computing the integral with $\hat{K}$ again one gets
\be
\label{cyF}
\begin{aligned}
\int_{\hat{\cal M}}& \underbrace{\,\hat K\wedge\dots \wedge \hat
K\,}_{7}\\ &=\int_{\cal M} \!D^{4} \bar D^{4} \Big( K_{m\bar n}
K_{p\bar q} K_{r\bar s} H_{\m\bar \m} H_{\n\bar \n} H_{\r\bar \r}
H_{\s\bar \s} \Big)\,\varepsilon^{\m\n\r\s} \varepsilon^{\bar
\m\bar \n\bar \r\bar \s} \de^{3}z \wedge \de^{3}\bar z
\end{aligned}
\ee where, by taking the eight superderivatives of the expression,
one finds several contributions due to the dependence of $K_{m\bar
n}$ and $H_{\m\bar \n}$ on $\,\theta$\,. It is easy to check that
there is always the conventional contribution. In the same way we
can compute the other integral and get \be\label{cyG}
\begin{aligned}
\int_{\cal \hat M}& \hat \varOmega \wedge \overline{\hat
\varOmega}\\&= \int_{{\cal M}^{3}}\!\Big(D^{4} \varOmega_{mnp\,
\m\n\r\s} \Big) \Big(\bar D^{4} \bar\varOmega_{\bar m\bar n\bar
p\,\bar \m\bar \n\bar \r\bar \s} \Big) \,\varepsilon^{\m\n\r\s}
\varepsilon^{\bar \m\bar \n\bar \r\bar \s} \de^{3}z \wedge
\de^{3}\bar z
\end{aligned}
\ee where we have used the fact that $\varOmega$ is holomorphic
both the in fermionic and bosonic coordinates. To take the
derivatives one has simply to pick out the coefficient
$\,\theta^{4}\,$ of the superfield $\varOmega_{mnp\,\m\n\r\s}$ and
there is no interference between the two factors.  

%If we assume
%that $ \varOmega_{mnp\, \m\n\r\s} $ is factorizable into a bosonic
%part and into a fermionic part, and $\varOmega_{\m\n\r\s} = \tau\,
%\theta_{\m} \dots \theta_{\s}$ where $\tau$ is function on ${\cal
%M}$, we find that
%\be
%\label{cyGA} \int_{{\cal \hat M}} \hat \varOmega \wedge
%\overline{\hat \varOmega} = \int_{{\cal M}}  |\tau|^{2}
%\,\,\varOmega \wedge \bar\varOmega\,. \ee In general $\varOmega$
%fails to be holomorphic and it satisfies a more general equation
%of the form
%\be
%\label{cyGB} \de \varOmega = F\,, \ee where $F_{4}$ has indices in
%the holomorphic and in the antiholomorphic sector, namely $F$ is a
%$(3,1)$ form. However, if also $\tau$ fails to be holomorphic
%then, by introducing $\varOmega' = \tau\, \varOmega$ and by
%requiring that
%\be
%\label{cyGC} \ln \tau = - \int (\varOmega^{-1})^{mnp} F_{mnp \bar
%q} \,\de z^{\bar q}\,, \ee one finds that $\de \varOmega' =0$ and
%\be
%\label{cyGD} \int_{\cal \hat M} \hat \varOmega \wedge
%\overline{\hat \varOmega} = \int_{{\cal M}} \left|e^{-\int
%(\varOmega^{-1})^{mnp} F_{mnp \bar q} \,\de z^{\bar q}
%}\right|^{2} \,\varOmega \wedge \bar\varOmega\,, \ee which depends
%on the field $F$\,.

Let us consider a further aspect. As we know very well from string
theory compactified on CY, the moduli of the space are
parametrized by the cohomology groups $\mathcal{H}^{(1,1)}$ and
$\mathcal{H}^{(2,1)}$. The first set parametrize the K\"ahler
deformations of the space and it is customary to write them in
terms of a basis $\omega_{\a}$ where $\a=1, \dots, h^{(1,1)}$.
Using this basis one finds the Yukawa coupling constants by
computing he intersection numbers $\,d_{\a\b\gamma}\,$ of the CY
given by the integral $\int_{{\cal M}} \omega_{\a}\wedge
\omega_{\b}\wedge \omega_{\g} = d_{\a\b\g}$\,. By embedding the CY
into a supermanifold, we can export the moduli $\omega_{\a}$ to
the new environment as above by adding the $\d$-function  terms.
Therefore, by computing the integral in the supermanifold we find
the same intersection numbers. To deviate from this situation, we
now propose a more radical extension of $\omega_{\a}$ by adding
the dependence on the fermionic coordinates and by modifying the
$\delta$-piece. Finally, the modified intersection numbers
$\,\tilde{d}_{\a\b\g}\,$ are given by \be\label{newint}
\int_{{\cal M}^{3}}\! D^{4} \bar D^{4} \Big( \omega_{\a m\bar n}\,
\omega_{\b p\bar q}\, \omega_{\g r\bar s} \,H_{\m\bar \m}
H_{\n\bar \n} H_{\r\bar \r} H_{\s\bar \s}
\Big)\,\varepsilon^{\m\n\r\s} \varepsilon^{\bar \m\bar\n\bar
\r\bar\s }\, \de^{3}z \wedge \de^{3}\bar z \ee where we have a
dependence upon the background fields $H_{\m\bar \n}$ because of
the seven powers needed in order to saturate the total degree.

These are simple examples to illustrate how the bosonic geometry
is influenced by extending the original CY space to a SCY. In the
same way one can use these equations to define new geometrical
quantities for a non-CY space. Namely one can start from a
K\"ahler non-CY space, then add fermionic coordinates such that
the complete supermanifold turns to be a SCY 
and finally, use equations \eqref{cyF},
\eqref{cyG}, \eqref{newint} to define new geometrical quantities
for the original space. The goal would be to see whether these
integrals can be evaluated by means of topological strings as
topological invariants. More detailed considerations will be left
to forthcoming publications.

As a final remark, we
want to suggest one of the most important application of our results: based
on the previous work \cite{Grassi:2006cd} one can assign a given supermanifold
on a bosonic manifold in such a way that the resulting manifold is super-Calabi-Yau.
Then, one can define a topological string on it and performs a mirror duality. Reducing the
amplitudes to the bosonic submanifold (known also as the body of the supermanifold) that
it would be a (super)-mirror pair of the original manifold. It would be very interesting to describe
the new manifold in terms of the original one.

\section*{Acknowledgments}

We thank G. Policastro, M Debernardi, D. Matessi, R. Catenacci and P. Fr\'e for 
useful discussions and comments.  

%%%%%%%%%%%%%%%%%%%%%%%%%%%%% References %%%%%%%%%%%%%%%%%%%%%%%%%%%%%%%%%%%%%%%%%%%%%%

\references

\end{document}